\documentclass[12pt,peerreviewca,final]{IEEEtran}
\usepackage{multirow}
\usepackage{subfigure}
\usepackage[pdftex]{graphicx}
\usepackage{array}
\DeclareGraphicsExtensions{.pdf,.gif}
\usepackage[center]{caption}

\renewcommand{\baselinestretch}{1.9}
\begin{document}

\title{The Dynamics of Vehicular Networks in Urban Environments}

\author{\IEEEauthorblockN{Nicholas Loulloudes George Pallis Marios D.
Dikaiakos\\}
\IEEEauthorblockA{Department of Computer Science, University of Cyprus}}

\maketitle

\begin{abstract}
Vehicular Ad hoc NETworks (VANETs) have emerged as a platform to support
intelligent inter-vehicle communication and improve traffic safety and
performance. The road-constrained, high mobility of vehicles, their unbounded
power source, and the emergence of roadside wireless infrastructures make VANETs
a challenging research topic. A key to the development of protocols for
inter-vehicle communication and services lies in the knowledge of the
topological characteristics of the VANET communication graph. This paper
explores the dynamics of VANETs in urban environments and investigates the
impact of these findings in the design of VANET routing protocols. Using both
real and realistic mobility traces, we study the networking shape of VANETs
under different transmission and market penetration ranges. Given that a number
of RSUs have to be deployed for disseminating information to vehicles in an
urban area, we also study their impact on vehicular connectivity. Through
extensive simulations we investigate the performance of VANET routing protocols
by exploiting the knowledge of VANET graphs analysis.
\end{abstract}

\section{Introduction}

Inter-vehicle communication (IVC) has emerged as a promising field of research
and development, where advances in wireless and mobile ad-hoc networks, global
positioning systems and sensor technologies can be collectively applied to
vehicles and result to great market potential. The idea of employing wireless
communications in vehicles dates back to the '80s, but recently the resolution
of governments and national traffic administrations to allocate wireless
spectrum for vehicular communications, along with the adoption of standards like
the Dedicated Short Range Communications (DSRC), has provided a real thrust in
the field of IVC or Vehicular Ad-Hoc Networks (VANETs\footnote{Although the
terms IVC and VANET are not identical, in this work we use the latter in order
to emphasize the ad hoc nature of these wireless networks.}). VANETs comprise
vehicle-to-vehicle and vehicle-to-infrastructure communications based on
wireless local area network technologies. Unlike cellular communication
networks, VANETs do not necessarily need continuous coverage, rather, they can
be supported by hot spots in correspondence of roadside infrastructure units
(RSUs), which aim to gap intermittent connectivity among vehicles.

The development of VANETs and their use in the deployment of vehicular
applications and services has been the target of research investigations in the
recent literature, as well as of industrial projects run by large
government-sponsored consortia (Vehicle Safety Consortium - USA, Car-2-Car
Communication Consortium and CVIS - EU), along with field trials (Vehicle
Infrastructure Integration, USA)~\cite{Car2Car, CVIS, VII}. Due to the nature of
vehicular mobility, VANETs are characterized by highly dynamic topologies and
are frequently prone to network disconnection and fragmentation. Consequently,
these inherent characteristics are bound to degrade the quality of services
provided by the VANET infrastructures. Therefore, the establishment of robust
VANETs that could effectively support multi-hop communications and applications
on a large geographical scale remains an open challenge.

\subsection{Motivation}

The motivation of this work stems from the fact that the study of the structural
properties of large, real-world, dynamic graphs (such as the Internet topology,
Web and e-mail graphs, collaboration, biological and on-line social networks)
has lead to crucial observations having significant influence in computer
science. For instance, the discovery of power laws in the Internet topology by
M. Faloutsos, P. Faloutsos and C. Faloutsos in~\cite{Faloutsos1999}, brought a
revolution to the field, since it enabled the design and the performance
analysis of information routing protocols. Therefore, the study of the
networking shape of VANETs is of paramount importance, currently gaining
considerable research attention, momentum, and expected to be of increasing
interest to the networking community in general.  With a better understanding of
network topology characteristics, the network researchers are able to design
network protocols which exploit the unique characteristics of VANETs.

Essentially, a vehicular network is a very challenging and dynamic environment
since it combines a fixed infrastructure (RSUs, e.g., proxies), and ad hoc
communications among vehicles. Despite the fact that it presents similarities
with the traditional Mobile Ad Hoc NETworks (MANETs), the mobile nodes in a
VANET (i.e. vehicles) are not energy-starving, are highly mobile, and their
mobility is constrained by the underlying road network topology. Moreover, the
existence of roadside infrastructure creates opportunities for optimized
communications such as data sharing and increased throughput while downloading
content on the move. Finally, the connectivity of VANETs is influenced by the
market penetration of communication equipment~\cite{Bai2009}. Apart from the
networking aspects, the applications which are expected to run over a VANET make
it also a unique environment: safety applications (accident avoidance near
intersections, speed ``regulation" for road congestion avoidance), peer-to-peer
music sharing, Internet access, they all pose interesting questions related to
protocol design and network deployment. The challenges featured by VANETs are
more related to the ones typically found in DTN (Delay Tolerant Networks) than
in infrastructure-based wireless networks.



During the process of designing and deploying a VANET, various questions must be
answered that pertain to protocol performance and usefulness. For instance, when
performing  message routing, a key question is ``which are the highest-quality
nodes (vehicles)?" \cite{Erramilli2008} to be trusted with the forwarding
process; when performing multi-hop geocasting, the question is how can we spread
the emergency messages with the minimal number of rebroadcasts so as to reduce
latency and packet collisions?; when performing periodic single-hop
broadcasting, how can we increase the average packet delivery?; when
provisioning for the placement of RSUs~\cite{Trullols2010}, in order to reduce
the average path length between the vehicles and the access points, what is the
best deployment strategy that maximizes the dissemination of information?;  when
designing mobility models~\cite{Musolesi2007}, which is the distribution of
``synapses" per node, i.e., whether there are any clusters (communities)?;
furthermore, when the network is disconnected, a significant question concerns
the identification of bridge nodes~\cite{Daly2007} which can ultimately be
en-charged with the ferrying of messages.

All these questions and many more require knowledge of the topological
characteristics of the VANET communication graph, where vehicles correspond to
vertices and communication links to edges. Specifically, vehicular mobility
affects the evolution of VANET connectivity over space and time in such a way
that critically determines the performance of networking protocols. Therefore, a
key question in vehicular networking is: ``\emph{which effects does mobility
generate on the vehicular topology and how can we exploit them in order to
improve vehicular communication protocols, especially for ``killer
applications", like large data files transfer and real-time information
dissemination~\cite{Nzouonta2009}?}".

\subsection{Paper's Contributions}
The objective of this work is twofold: a) to provide a ``higher order" knowledge
of the time-evolving topological characteristics of the VANET communication
graph, as compared to the ``first-order" knowledge provided by the studies
reported in~\cite{Pallis2009,Fiore2008,Srivastava2008,Viriyasitavat2011} and b)
to investigate how to use this knowledge towards improving the performance of
VANET routing protocols. The present work continues and improves upon the
authors' preliminary efforts in~\cite{Pallis2009} towards deriving insightful
implications for the vehicular networking community. Using both real and
realistic vehicular mobility traces, as well as current and future wireless
communication technologies, in a range of real-world urban environments, we
study the structure and evolution of the communication graphs under different
market penetration ratios of VANET-enabled vehicles. Also, we study the impact
of the presence of stationary RSUs has on the VANET connectivity, since such an
infrastructure has a potentially significant effect on inter-vehicle
connectivity in urban environments.  Our research goes one step further and we
examine the impact of our findings in the design of VANET routing protocols.
Through extensive simulations, we explore how the knowledge of the networking
shape of vehicular network could enhance the design of routing protocols. To the
best of our knowledge, this is the first attempt to study the network
connectivity of urban traffic as well as its impact in the design of  VANET
routing protocols in such a systematic and comprehensive way. The main
contributions of this work can be summarized as follows:

\begin{itemize}
\setlength{\itemsep}{-3pt}
\item \emph{Data Knowledge Perspective:} A thorough study of the visible and
``latent" structure of the vehicular network, including metrics used in earlier
studies~\cite{Fiore2008,Pallis2009,Srivastava2008}, as well as several other
metrics traditionally used in the field of social network analysis, i.e.,
community analysis. Using both real and realistic mobility traces, we study the
networking shape of VANETs in urban environments under different transmission
and market penetration ranges. Given that a number of RSUs have to be deployed
for disseminating information to vehicles in an urban area, we also study their
impact on vehicular connectivity.
\item \emph{Engineering Perspective:} Based on the analysis conducted, we
examine the implications of our findings in the design of two indicative VANET
routing protocols (VADD~\cite{Zhao2006}, GPCR~\cite{Lochert2005}). Through
extensive simulations, we investigate how the performance of routing protocols
can be improved using the knowledge of VANET graph analysis. We provide
significant perspectives and insights into how vehicular protocols should be
designed for urban traffic.
\end{itemize}

The rest of the paper is organized as follows: Section~\ref{sec-relevant-work}
briefly surveys the relevant work; Section~\ref{sec-metrics} describes the
metrics used in the present study to characterize the evolution of VANET
communication graph; Section~\ref{sec:Data} provides detailed information
concerning the source of the data studied here. Section~\ref{sec:Observations}
records the findings of the study. Section~\ref{sec:simulation} examines whether
the findings from the VANET graph analysis can be utilized by two
known VANET protocols (VADD and GPCR) during their routing decision processes.
Finally, Section~\ref{sec:Conclusions} concludes the article.

\section{Relevant Work}
\label{sec-relevant-work}

Our work draws inspiration from the rich body of prior work on frameworks for
studying the temporal evolution of several real graphs~\cite{Leskovec2007}.
These graphs arise in a wide range of  domains (i.e. autonomous systems, e-mail
networks, citations) and their study leads to significant implications since
most of real-world dynamic networks (online social networks, Internet etc.) have
been proved to follow some topological statistical features (i.e. features of
scale-free networks, small-world properties, power-law degree distribution
etc.).

In this work, we focus on exploring the time-evolving VANET graphs where the
mobility of nodes affects the evolution of network connectivity over space and
time in a unique way. Inter-vehicle communication (IVC) has emerged as a
promising field of research~\cite{Dressler2011,Kosch2006, Nathan2006}, where
advances in wireless and mobile ad-hoc networks can be applied to real-life
problems (traffic jams~\cite{Nadeem2004,AutoNet,SmarterTraffic}, road
accidents~\cite{PATH, PReVENT} etc) and lead to a great market
potential~\cite{ITS-Japan,CVIS}. Connectivity dynamics, in turn, determine the
performance of networking protocols, when they are employed in vehicle-based,
large-scale communication systems. The value of the connectivity analysis of ad
hoc networks is so fundamental that recently a competition-experiment has been
started --- the MANIAC experiment~\cite{Srivastava2008}--- to study network
connectivity, diameter, node degree distribution, clustering, frequency of
topology changes, route length distribution, route asymmetry, frequency of route
changes, and packet delivery ratio. The obtained results show a high degree of
topology and route changes, even when mobility is low, and a prevalence of
asymmetric routes, both of which contradict assumptions commonly made in MANET
simulation studies. Regarding link duration statistics, authors
in~\cite{Sadagopan2003} used simulations to study the probability densities of
link lifetime and route lifetime for some mobility models. According to this
study, the path duration seems to be a good metric in order to predict the
general trends in the performance of vehicular routing protocols. Another sound
observation is that they showed the relationship between the path duration and
other critical parameters such as  the transmission range and the average
relative speed of the mobile nodes, and the average number of hops in the path.
Also, well-known concepts from social network analysis have been used as
primitives to design advanced protocols for routing and caching in DTNs and ad
hoc sensor networks. In~\cite{Daly2007}, the betweenness centrality index and
its combination with a similarity metric (comprising both the SimBet metric)
have been used to select forwarding nodes to support information routing in
DTNs. Results showed that data dissemination is improved if the messages are
delivered through nodes which have high SimBet utility values. The betweenness
centrality has also been used in~\cite{Dimokas2008} to design a cooperative
caching protocol for wireless multimedia sensor networks. This protocol selects
the mediator nodes that coordinate the caching decisions based on their
``significant" position in the network. Likewise, the MaxProp
protocol~\cite{Burgess2006} transfers messages based on the mobility of
intermediate nodes. In a related study~\cite{Ghandeharizadeh2006}, the authors
combine short-range communication and cellular communication to facilitate query
processing in VANETs. Yoneki studied the impact of connective information
(clustering, network transitivity, and strong community structure) on epidemic
routing in a series of works~\cite{Yoneki2008}.


In the context of vehicular networking, authors of~\cite{Conceicao2008} present
a preliminary characterization of the connectivity of a VANET operating in an
urban environment. They transform the vehicular network into a transitive
closure graph. Then, the temporal evolution of the average node degree is
presented. Nonetheless, the authors do not perform a deep analysis of the
networking shape of vehicular mobility and limit their study only to the average
node degree for a small time interval. In~\cite{Seada2008}, the authors set up a
real-world experiment consisting of~$10$ vehicles making loops in a $5$-mile
segment of a freeway. They focus on the connectivity issues without
investigating the topological properties of the VANET graph. Authors
in~\cite{Fiore2008} study the node degree distribution, link duration,
clustering coefficient and number of clusters for VANET graphs under various
vehicular mobility models. The objective of~\cite{Fiore2008} focuses on studying
the topological properties of different mobility models and explaining why
different models lead to dissimilar network protocol performance. The authors
of~\cite{Kafsi2008} provide an analysis of the connectivity of vehicular
networks by leveraging on well-known results of percolation theory. Using a
simulation model, they study the influence of vehicle density, the proportion of
equipped vehicles, transmission range, traffic lights and roadside units.
Similarly, authors in~\cite{Meireles2009} study the distributions of node degree
and link duration in VANETs using a realistic urban traffic simulator. In a
recent study~\cite{Wang-Hei2011}, authors study the vehicular network in order
to develop a stochastic traffic model for VANETs. This model captures spatial
and temporal characteristics of a vehicular network, vehicle movement, link
condition, and node connectivity. There has also been a recent work
in~\cite{Viriyasitavat2011} which is closely related to ours. Specifically, the
authors present a comprehensive analytical framework, as well as a simulation
framework, for network connectivity of urban VANETs, using some key system
parameters such as link duration, connection duration, and rehealing time. The
analytical framework leads to closed-form expressions which capture the impact
of four critical parameters (network density, transmission range, traffic light
mechanisms, and size of a road block) on network connectivity. Unlike our work,
the authors do not study how the observations derived from the study of the
structure of the vehicular network affects  well known vehicular communication
protocols (e.g., VADD~\cite{Zhao2006}, GPCR~\cite{Lochert2005}). In addition, we
study the impact of roadside units in the networking shape of VANET graphs.

In~\cite{Pallis2009}, we studied the structure and evolution of a VANET
communication graph using realistic traces from the city of Zurich. However, the
traces utilized in~\cite{Pallis2009} are characterized by highly variable
vehicular densities. In this work, we extend the analysis of VANET graphs from
the previous studies in~\cite{Fiore2008,Kafsi2008,Viriyasitavat2011} by
introducing additional new measurements and new data sets. Considering that the
usefulness of the findings of this research study depends on the realism and
completeness of the data upon which the study is based, we use both real and
realistic vehicular traces. Another extension of this work is that we examine
the implications of our findings in the design of two indicative VANET routing
protocols. Through extensive simulations, we study how the knowledge of the
networking shape of vehicular network can be beneficial.

\section{Graph Metrics Examined}
\label{sec-metrics}

\begin{figure}[!hbt]
\centering
\includegraphics[width=3in]{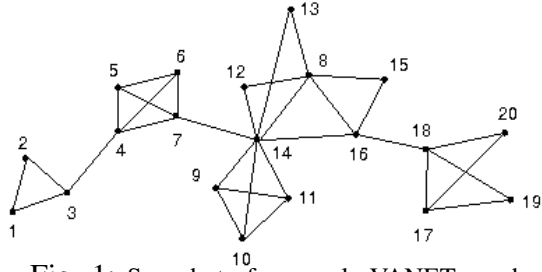}
\vspace*{-\baselineskip}
\caption{\small Snapshot of a sample VANET graph.}
\label{sample-graph}
\vspace{-6mm}
\end{figure}

This section contains the definitions of the metrics used in the study. We
categorize the examined metrics as {\it network-oriented} , {\it centrality},
{\it link duration} and {\it cluster-oriented}. All node IDs mentioned in this
section refer to the sample graph of Figure~\ref{sample-graph}. For the sequel,
we will consider~$G(t)$ to be an undirected graph of VANET at time~$t$, where
vehicles correspond to the set of vertices~$V(t)=\{u_i\}$ and communication
links to the set of edges~$E(t)=\{e_{ij}\}$. An edge~$e_{ij}(t)$ exists,
if~$u_i$ can communicate directly with~$u_j$ at time~$t$, with~$i\neq j$.

\subsection{Network-oriented metrics}
Network-oriented metrics depict the shape of VANETs, capturing and
quantifying the richness of network connectivity. These metrics reflect the
topological properties of a vehicular network, stimulating interesting
considerations on how network protocols could take advantage of vehicular
mobility to improve their performance.

\begin{itemize}
\setlength{\itemsep}{-3pt}
\item \textbf{Node degree}. The number of vehicles within the
  transmission range of a node. Formally, the degree of~$u_i$ at
  time~$t$ is defined as $D_{i}(t)=\|\{u_j \mid \exists
  e_{ij}(t)\}\|$.

\item \textbf{Effective Diameter}. The minimum distance in which the 90th
percentile of all connected pairs of vehicles can communicate with each other. It
is a smoothed form of network diameter which we use for our studies.
\item \textbf{Density}. Defined as the ratio between the number of edges in the
$G(t)$ and the maximum number of edges possible for $G(t)$.
\end{itemize}

\subsection{Centrality metrics}
Centrality metrics have been developed in social network analysis to quantify how
important particular individuals are in social networks. The objective is to find
people who are central to communication and important for information
dissemination. Centrality metrics are designed such that the highest value
indicates the most central node. In~\cite{Pallis2009}, we observed several
centrality metrics and concluded that for VANETs, Betweenness Centrality and the
Lobby index provide a good measure as they relate to the expected role a node
plays within the vehicular ad hoc network.

\begin{itemize}
\setlength{\itemsep}{-2pt}
\item \textbf{Betweenness Centrality~\cite{Wasserman1994}}. Defined as the
fraction of the shortest paths between any pair of nodes that pass through a
node. The betweenness centrality of a vehicle $u_i$ at time $t$ is:
   \begin{equation}
   BC_i(t)=\sum_{j\neq k}\frac{sp_{j,k}(u_i,t)}{sp_{jk}(t)}
   \end{equation}

\noindent where $sp_{jk}$ is the number of shortest paths linking vertices $j$
and $k$ at time $t$ and $sp_{j,k}(u_i,t)$ is the number of shortest paths linking
vertices $j$ and $k$ that pass through $u_i$ at time $t$. Betweenness centrality
is a measure of the extent to which a vehicle has control over information
flowing between others (e.g., $BC_{14}= 0.668$).

\item \textbf{Lobby  Index~\cite{Korn2008}}. The lobby index of a given vehicle
$u_i$ at time $t$, denoted as $L_{i}(t)$, is the largest integer~$k$ such that
the number of one-hop neighbors of~$u_i$ in graph~$G(t)$ with degree at least~$k$
equals~$k$. This metric can be seen as a generalization of $D_{i}(t)$, conveying
information about the neighbors of the node as well (e.g., node with ID~$8$ has
lobby index~$2$).

\end{itemize}

\subsection{Link level metrics}
The metrics of this category reflect the network connectivity over a period of
time. These metrics are critical both for the periodic single-hop broadcast
primitive, where a vehicle broadcasts its current information (location,
velocity) in its direct neighborhood, and also for the multi-hop geocast
primitive, where several data exchanges between vehicles take place during one
communication session.

\begin{itemize}
 \item \textbf{Number of connected periods}. The number of established links
 between a pair of vehicles within a given time period. A connected period is the
 continuous time interval during which a physical link is established between two
 vehicles as a consequence of one being in the transmission range of the other.

  \item \textbf{Link duration}. The time duration of a connected period.
  Formally, the duration $l_{ij}(t)$ of the link from~$u_i$ to~$u_j$ at time~$t$
  is defined as $l_{ij}(t)=t_{c}-t_{o}$, if  $\exists e_{ij}(t)$, where $t\in
  [t_o,t_c]$ and $\not\exists e_{ij}(t')$, where $t'<t_{o}$ or $t'>t_{c}$.


 \item \textbf{Re-healing time}. The time span between two successive connected
 periods of a pair of vehicles. Formally, the duration $r_{ij}(t)$ of the
link from~$u_i$ to~$u_j$ at time~$t$ is defined as $r_{ij}(t)=t_{m}-t_{k}$, if
$\exists e_{ij}(t)$, where $t\in  [t_a,t_k]$ and $\exists e_{ij}(t)$, where $t\in
 [t_m,t_z]$ and $t_a<t_k<t_m<t_z$.

\end{itemize}

\subsection{Cluster metrics}
The examined cluster metrics present the dynamic properties of clusters and
dense subgraphs established inside a VANET communication graph. The existence of
vehicle communities is important in terms of information propagation, since they
can act as ``data islands'' in a vehicular environment.

\begin{itemize}
\setlength{\itemsep}{-2pt}
\item \textbf{Number of Clusters}. The number of co-existing, non-connected
clusters of nodes at a given instant. We define as cluster a connected group of
vehicles. A connected group is a sub-graph of the network such that there is a
path between any pair of nodes.

\item \textbf{Clustering Coefficient}. It measures the cliquishness of a network.
The clustering coefficient $p_{k}(t)$ of a cluster~$k$ at time~$t$ (as defined
in~\cite{Fiore2008}) is:
   \begin{equation}
    p_{k}(t)=\frac{2|E_{k}(t)|}{|N^k(t)|(|N^k(t)|-1)},
   \end{equation}
   \noindent
   where $|E_{k}(t)|$ is the number of existing links in cluster $k$ at time $t$
   and $|N^k(t)|$ is the number of nodes in cluster $k$ at time $t$. The
   clustering coefficient has a maximum value~$1$ if the cluster is a clique.

\item \textbf{Number of Communities~\cite{Newman2004}}. The number of existing
communities at a given instant. A community is defined as a dense sub-graph
where the number of intra-community edges is larger than the number of
inter-community edges. In order to identify communities, we transform $G(t)$ to
directed graph so as $D_{U_{i}}^{in}(t)=D_{U_{i}}^{out}(t)=D_{U_{i}}(t)$, where
$D_{U_{i}}^{in}(t), D_{U_{i}}^{out}(t)$ is the in-degree and out-degree of
node~$u_i$ at time $t$. Formally, a sub-graph $U(t)$ of a VANET graph $G(t)$ at
time $t$ constitutes a community, if it satisfies:
  \begin{equation}
  \label{generalcom}
   \sum_{u_{i} \in U}(D_{U_{i}}^{in}(t))(U(t)) > \sum_{u_{i} \in
   U}(D_{U_{i}}^{out}(t))(U(t)),
  \end{equation}
  i.e., the sum of all degrees within the community $U(t)$ is larger than the sum
  of all degrees toward the rest of graph the~$G(t)$.
  
\item \textbf{Community Modularity~\cite{Newman2004}}. Community modularity
quantifies in the range of -1 to 1 the division of the graph into communities.
Good divisions, which have high modularity values, give communities with dense
internal connections and weak connections between different communities. The
community modularity $Q$ of a network at time~$t$ is:

\begin{equation}
\label{modularity}
	Q_{t}=\frac{1}{2m}\sum_{U_{i}U_{j}}[A_{ij}-\frac{D_{U_{i}}D_{U_{j}}}{2m}]\delta(U_{i},U_{j})
\end{equation}

where $A_ij$ is an element of the adjacency matric of the network, $m$ number of
links and $D_{U_{i}}$ the degree of node $i$. 


\end{itemize} 
\section{Traffic Data Studied}
\label{sec:Data}

This section provides an in-depth overview of the dataset utilized in the study
of the VANET communication graph.

\subsection{Vehicular Mobility Traces}
The structure and evolution of the VANET communication graph is dictated by
vehicles mobility patterns. Hence the utilization of both real and realistic
mobility traces in this study was of outmost importance. Except of real traces,
we also study traces derived from mobility models which are proven to be
realistic. For both real and realistic vehicular traces, we consider real-world
urban environments.

\subsubsection{Real Traces}
Real mobility traces were obtained through the Smart City Research Group,
~\cite{SCRG}. SCRG maintains a dataset of real GPS traces collected from taxis
traveling throughout Shanghai, China, in a 24-hour time period. We developed a
tool in order to parse the taxi traces and converted them from GPS cylindrical
coordinates to Cartesian plane coordinates for better manipulation. A $2Km
\times 2Km$ rectangular region around Shanghai city-center was isolated, and the
mobility traces of all the taxis that run within this region for a 2-hour time
period, were utilized. This clipping process resulted in obtaining the mobility
traces of 704 distinct vehicles for our study. According to the authors'
knowledge, the SCRG data set is the largest, publicly available dataset for
vehicular traces.


\subsubsection{Realistic Traces}
Due to the size constraints of real traces, we enhance our evaluation by
studying realistic mobility traces in real urban road topologies. We employ
vehicular traces that follow a steady state behavior, where the number of
vehicles in a given region remains constant over time. To this end, realistic
mobility traces were generated using the VanetMobiSim~\cite{VanetMobiSim09}
vehicular mobility generator. By employing known traffic generation models,
VanetMobiSim outputs detailed mobility traces over real-world, accurate, city
maps available in the Topological Integrated Geographic Encoding and Referencing
(TIGER) database~\cite{TigerMaps} from U.S. Census Bureau. The realism of the
mobility models utilized in VanetMobiSim has been validated extensively
in~\cite{VanetMobiSim09} using benchmark tests from vehicular traffic flow
theory which are highly accepted by the transportation community.

In particular, we chose to generate vehicular traffic within a $2Km \times 2Km$
area bounding the city center of Los Angeles, CA. In the generated scenario all
vehicles were set to follow the Intelligent Driver Model with Lane Changes
(IDM-LC). More specifically, IDM-LC falls into the car-following mobility models
category since individual vehicle behaviour depends on the behaviour of the
preceding vehicle. It extends the basic IDM model, in the sense that it adds
intersection handling capabilities to the behaviour of vehicles and the
possibility that vehicles change lanes and overtake other preceding vehicles. By
examining the respective TIGER map, VanetMobiSim was set to synchronize the
operation of traffic signals of 110 different intersections.

To achieve an even higher level of realism in the vehicular traffic generated,
we opted to define the average vehicle density (and hence the total number of
vehicles) for the area under study, by examining the 2010 Annual Average Daily
Traffic (AADT) statistics, which are publicly available through California's
Department of Transportation Traffic Data Branch~\cite{CalTrafficDataBranch}.
AADT measures the total volume of vehicle traffic on a given road for a year
divided by 365 days. In general, AADT is obtained through traffic counting using
electronic equipment installed on the roadside. Our study on a large number of
roads in the area of interest, indicated an average density of~\textit{10
veh/lane/Km} during normal driving conditions (not peak hours). With
approximately 700 Km of road length (one-way and number of lanes for each
individual roads included), VanetMobiSim was set to generate vehicular traffic
for a 2 hour period, constantly maintaining 7000 vehicles in the area of
interest.


\subsection{Wireless Communication Technologies}
Another key factor that undoubtedly determines the structure and evolution of
the VANET communication graph is the technology through which vehicle-to-vehicle
(V2V) and vehicle-to-infrastructure (V2I) wireless communication is achieved.
Currently, the de-facto wireless communication protocol used in research works
(simulation, as well as field studies) in the VANET literature is the normal
IEEE 802.11a WiFi. However, since this protocol has not been designed having
mobility in mind, an amendment to the IEEE 802.11 standard has been proposed to
add wireless access in vehicular environments (WAVE). This amendment, designated
as IEEE 802.11p, defines the necessary enhancements to support data exchange
between high-speed vehicles and between vehicles and roadside infrastructure.


The IEEE 802.11p protocol is still under development and only limited trial
implementations exist. To the best of our knowledge, the only available field
measurements for the IEEE 802.11p are the communication performance results
in~\cite{CVISResults} from the CVIS project~\footnote{CVIS technology is using
the CALM M5 ISO standard that incorporates the IEEE 802.11p PHY/MAC}. These
results indicate that, irrespective of vehicle speed, effective communication
can be achieved at a distance of 300m with a good and relative constant data
rate (~5 Mbit/s). Our extensive literature review, has indicated that the value
of 300m is also referenced in~\cite{YEYR2004} as the indicative V2V
communication distance using the IEEE 802.11p protocol.

Therefore, our study examines network graphs that capture wireless network
connections established between vehicles separated by a distance of at most 300
meters.

\subsection{Market Penetration}

The market penetration and the resultant background traffic has a significant
effect on the network connectivity of urban traffic since it affects the VANET
density. Thus, the spatio-temporal characteristics of the communication graph
exhibited in each of the aforementioned mobility scenarios were evaluated using
different market penetration values for VANET-enabled vehicles. We mainly study
the network graph for 5 different penetration ratios, starting with only 20\% of
the total number of vehicles being VANET-enabled and gradually reaching 100\% in
20-percentile increments. In some occasions we additionally study the network
graph behavior in much lower penetration rates of 1\%, 5\%, 10\% and 15\%.

\subsection{Stationary Road-Side Units}
In this work, we study extensively how the presence of RSUs influences the
structure and evolution of the VANET communication graph. To achieve this, we
extend the above realistic scenario in Los Angeles, by adding stationary RSUs.
Assuming, that RSUs run IEEE 802.11p in addition to any other protocol, their
effective communication distance in our evaluation was set to 300m. We utilize
the Wigle.net~\cite{Wigle} online catalog in order to extract the position of
Wi-Fi hot-spots in the area of study, which potentially could be utilized as
RSUs. Wigle maintains a publicly available and constantly user-updated catalog
of wireless networks throughout the US, with each network attributed with
identification (SSID), location (latitude/longitude), and security
(open/private) information.

We developed a tool to extract all open Wi-Fi Access Points (AP) in the area of
study and placed then on the map by converting the GPS cylindrical coordinates
to Cartesian plane coordinates. This process resulted in 427 APs that emulate
the existence of RSUs in our study.


\subsection{Testbed Overview}
Each of the above scenarios runs for 2 hours (7200 sec). We allow a 1000 second
warm-up period at the beginning before obtaining any measurements in order to
achieve some level of stability in the network. Consequently, we study snapshots
of the VANET communication graph taken every 1 second. In total, for the above 2
mobility scenarios (Shanghai and Los Angeles), the different penetration ratios
and the 1 complementary RSU scenarios (RSUs were not considered for real
traces), a total 167400 snapshots of the VANET communication graph were
recorded. Table~\ref{tbl:Setup} provides an overview of the evaluation testbed.

\begin{table}[t]
\centering
\tiny
\begin{tabular}{|l|c|c|c|}\hline
& \textbf{Shanghai}& \textbf{Los Angeles} \\\hline\hline
Trace Type & Real & Realistic \\\hline
Area Size & $4Km^2$ & $4Km^2$ \\\hline
Vehicle Density (veh/lane/Km) & n/a & 10 \\\hline
Average Vehicle Speed (m/s) & 4.72 & 3.12 \\\hline
Total Vehicles & 704 & 7000 (per sec)\\\hline\hline 
Vehicle Transmission Range & \multicolumn{2}{|c|}{ 300m (802.11p)}
\\\hline\hline 
RSUs & n/a & 427 \\\hline
RSU Transmission Range & n/a & 300m\\\hline
Penetration Ratio & \multicolumn{2}{|c|}{1\%, 5\%, 10\%, 15\%, 20\%, 40\%, 60\%,
80\%, 100\%}\\\hline
Simulation Time & \multicolumn{2}{|c|}{7200 sec (1000s warm-up)}\\\hline
Snapshots Interval & \multicolumn{2}{|c|}{every 1 second}\\\hline
\end{tabular}
\caption{Simulation setup parameters}
\label{tbl:Setup}
\vspace{-6mm}
\end{table}


\section{Observations}
\label{sec:Observations}

This section presents the findings of our study related to the laws governing
the networking shape of vehicular connectivity in both real and realistic
scenarios. Note that our work does not examine the dynamics of the VANET
communication graph under medium access problems (contention and
interference)~\cite{Jarupan2011}. Our work shows that, based on factors such as:
(i) mobility pattern of vehicles, (ii) a given transmission range, and (iii) the
existence of RSUs, there is a possibility that a number of communication links
could be established (if required) at a given point in time. The set of all such
possible links, dictate the characteristics of the VANET graph, and knowledge of
such information is crucial in the design/operation of routing and data
dissemination protocols. However, the particular of how these links (i.e the
shared medium) are accessed, is the work of MAC protocols, something which is
out-of-scope of our work.

In the interest of space and clarity, we present the analysis results for 1 hour
(3600 sec). The respective scenario and simulation parameters are indicated
accordingly. The data analysis was performed on a machine with an 8-core
Intel(R) Xeon(R) CPU @ 2.83GHz and 24GB RAM, using the aid of two graph/network
analysis frameworks, JUNG~\footnote{JUNG: Java Universal Network/Graph Framework
- http://jung.sourceforge.net (accessed August 2011)} and SNAP~\footnote{SNAP:
Stanford Network Analysis Platform - http://snap.stanford.edu  (accessed August
2011)}. The complete analysis duration of each individual graph varied between
5mins and 32hrs, depending on the selected penetration level.

\subsection{Network Analysis}
\label{subsec:NetworkAnalysis}

\begin{figure}[t]
    \centering
     \subfigure [Edges vs. Market Penetration Ratio]{
        \includegraphics[width=0.31\textwidth]{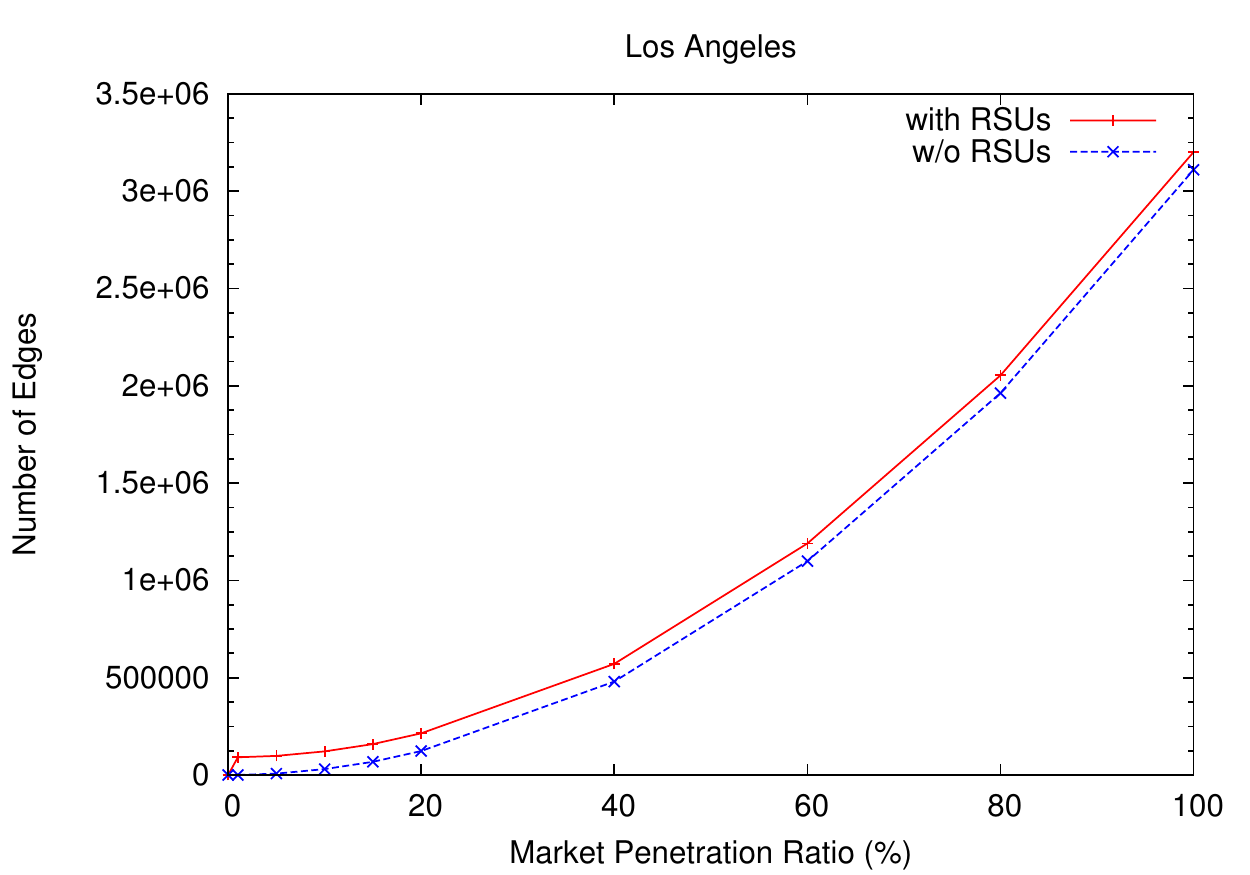}
        \label{fig:EdgesPratio}
   }
    \subfigure [Edges vs. Transmission Range (P=100\%)]{
        \includegraphics[width=0.31\textwidth]{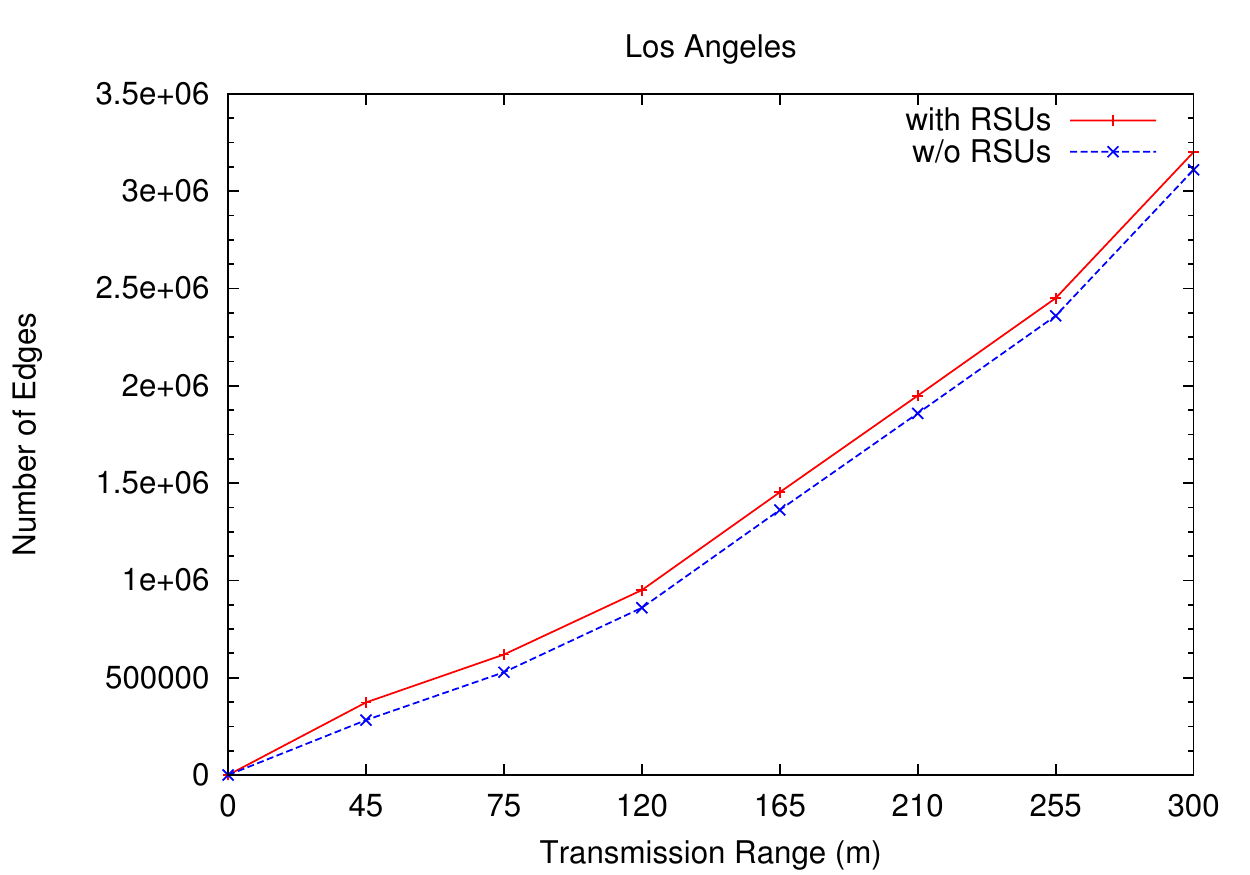}
        \label{fig:EdgesTx}
    }
     \subfigure[Edges vs Time]{
        \includegraphics[width=0.31\textwidth]{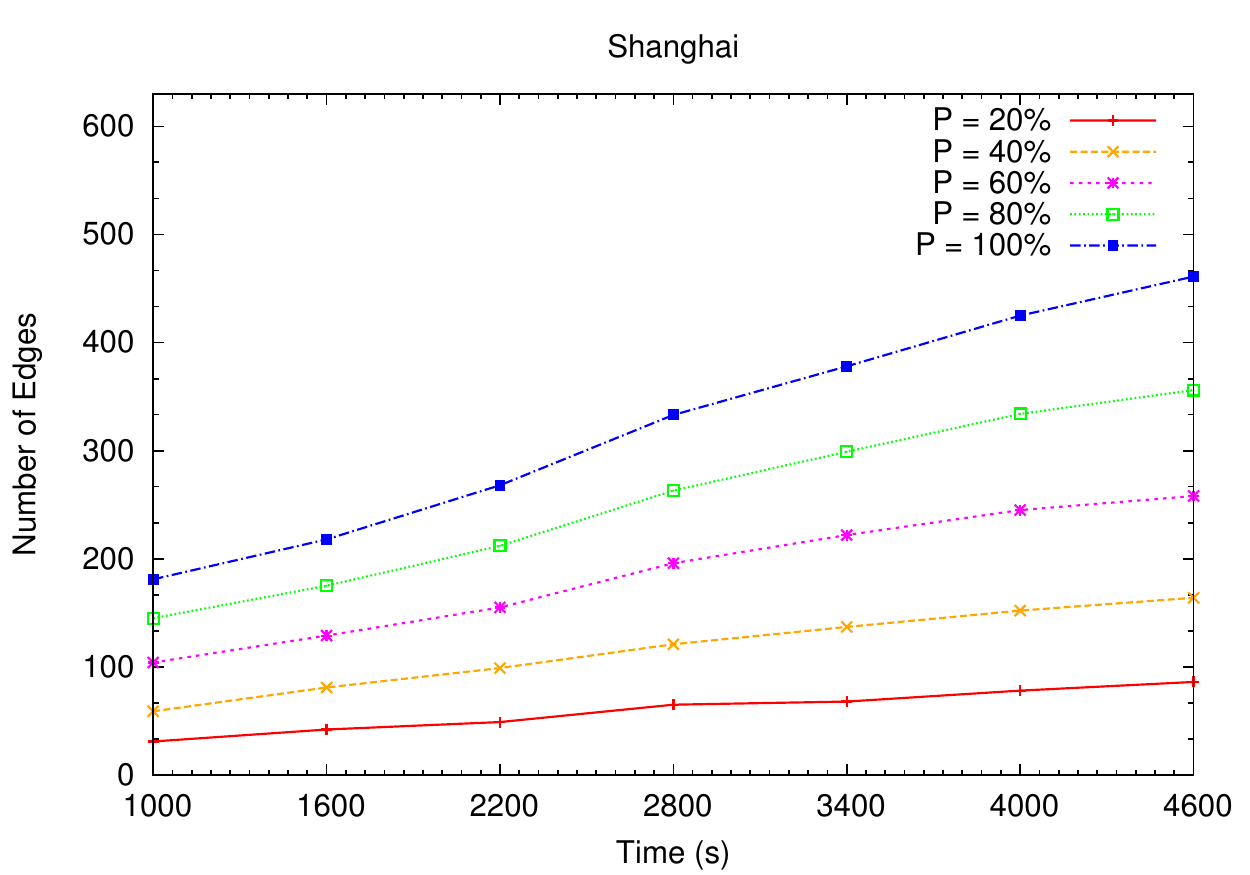}
        \label{fig:EdgesReal}
    } \caption[]{}
    \label{fig:edges}
    \vspace{-5mm}
\end{figure}

The spatio-temporal evolution of the VANET is primarily influenced by two
interrelated factors: i)~the number of vehicles and stationary RSUs that
participate in the network at any given time instance and, ii)~the effective
transmission range of the wireless communication hardware in use. Initially, to
gain elementary knowledge on this evolution, we study how the \emph{\textbf{size
of the VANET}} (number of edges and consequently the number of connections)
changes as a function of the above factors. To compensate for the fact that, the
realistic scenarios under study, follow a steady-state behaviour - that is the
number of vehicles remains constant over time - we opt to study the VANET
evolution as a function of different market penetration ratios and transmission
ranges. Figures~\ref{fig:EdgesPratio} and~\ref{fig:EdgesTx} illustrate the VANET
size over different market penetration ratios~\textit{(P)} and transmission
ranges~\textit{(Tx)} respectively for the Los Angeles. Specifically, we observe
the VANET graph is growing almost linearly as the penetration rate and
transmission ranges increase, an evident behaviour in all the examined scenarios
irrespective of time. Figure~\ref{fig:EdgesReal} illustrates the VANET size over
time in the real scenario (Shanghai), as more vehicles enter the map.

In agreement with the findings of~\cite{Leskovec2007}, we observe that
\textbf{\emph{average node degree}} increases as the VANET grows. As expected,
increasing the penetration ratio incrementally from as low as ~\emph{P}=1\%
to~\emph{P}=100\%, results in an increase of node degree(see
Table~\ref{table:RSUDegree}), however this rate of growth, decreases linearly.
This finding is quite important, since it provides the ability to estimate the
future size of a VANET, given forecasts of V2V and V2I technology penetration.
Such estimates allow engineers to plan-ahead the network capacity, as well as,
make fault tolerance and recovery provisions. It is worth noting that, despite
the small values of node degree observed in the Shanghai scenario (due to the
sparse nature of mobility traces), the node degree increase rate follows the
similar trend of the Los Angeles scenario.

\begin{figure}[t]
    \centering
     \subfigure[P=20\% (exemplary clusters circled)]{
        \includegraphics[width=0.30\textwidth]{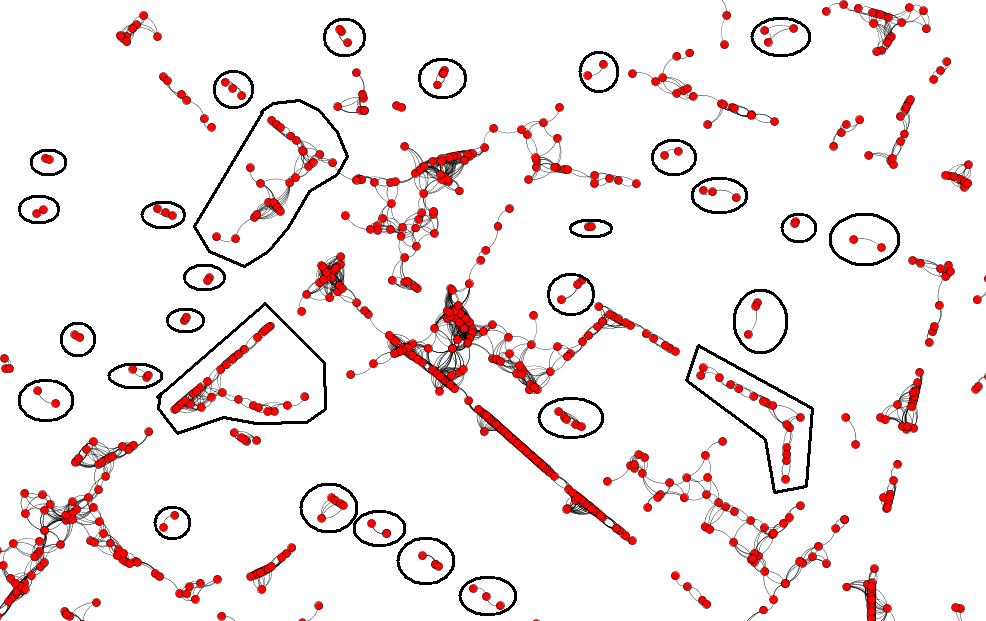}
    } \subfigure[P=100\%]{
        \includegraphics[width=0.30\textwidth]{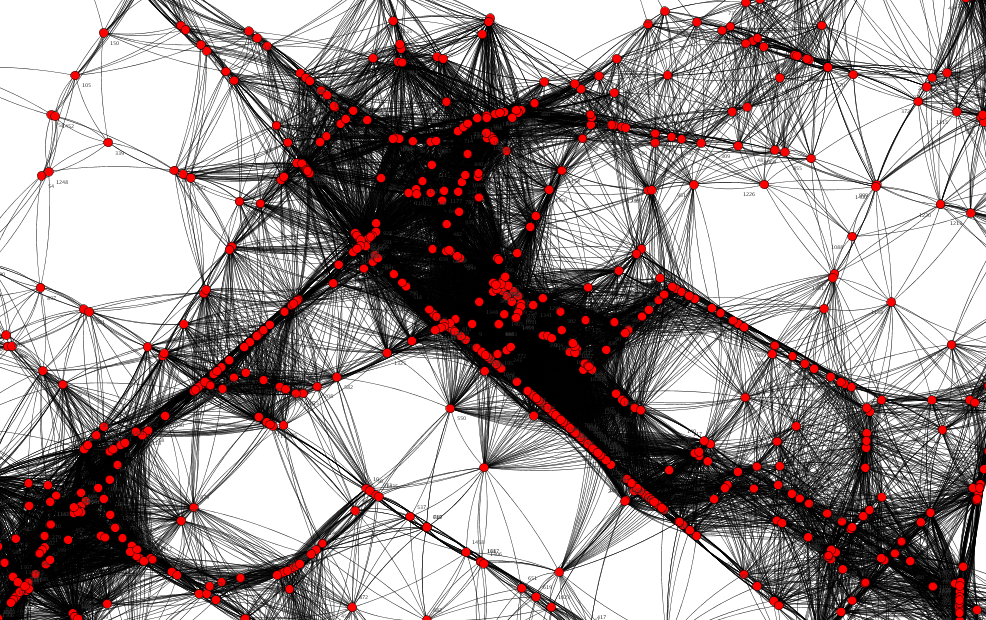}
    } \caption{Shanghai VANET snapshot w/o RSUs}
    \label{fig:VanetDump}
    \vspace{-5mm}
\end{figure}

A sound observation that occurs by examining the \emph{\textbf{degree
distribution}} of boths scenarios (figures not shown in the interest of space)
is that VANET graphs demonstrate a regularity that is unlikely to be a
coincidence. Specifically, we observe that at low penetration ratio, the degree
distribution of VANETs is skewed. As an example, for the Shanghai scenario and
$P<60\%$, VANET graphs have the following characteristics: the distribution
$P(k)$ of node degrees $k$ follows $P(k)\sim k^{-\gamma}$, with exponent
$\gamma$ often lying between 2 and 3. This indicates that low penetration rates
tend to fragment the VANET to several small clusters, while high penetration
rates favor the formation of a single large cluster and few smaller clusters at
its surrounding (cluster analysis follows in~\ref{subsec:ClusterAnalysis}). An
increase in the penetration ratio, improves the network connectivity
considerably, as each individual node is able to establish connections with even
more vehicles. Figure~\ref{fig:VanetDump}, obtained using~\cite{Spanakis2011},
provides a zoom-in snapshot on the shape of the VANET established in the
simulated area of Shanghai and gives a visual indication on the effects of
penetration rate in the formation of clusters.



Examining the degree distribution of the real scenario even further, we observe
that in the case of~\emph{P}=20\%, the average node degree remains significantly
low. Particularly, less than $50\%$ of the total vehicles are connected to more
than 4 neighbour vehicles. As a matter of fact, our study shows that these
vehicles are the ones located at the proximity of densely populated
intersections (4-way or higher intersections).


Our study shows that the introduction of \emph{\textbf{stationary RSUs}} in
cases of low penetration provides a significant potential to improve vehicle
connectivity. Evidently, RSUs improve inter-vehicle connectivity significantly
at low penetration ratios (~$\emph{P}<20\%$), in contrast to higher vehicle
densities where connectivity remains more or less constant. Considering the Los
Angeles scenario when~\emph{P}=10\% and after the deployment of 427 RSUs, the
median node degree is doubled. Table~\ref{table:RSUDegree} presents the node
degree statistics under the presence of RSUs and different penetration ratios.


\begin{table}[t] \tiny \centering \begin{tabular}{|l|c|c|c|c|}\hline
\multicolumn{5}{|c|}{\bf Los Angeles Area | Transmission Range = 300m} \\\hline
{Penetration} & {\bf Min}   & {\bf Max} & {\bf Mean} & {\bf Median}\\\hline\hline
1\% & 1 (1) & 426 (16) & 372.27 (8.10) & 426 (8) \\\hline
5\% & 1 (1) & 426 (90) & 258.34 (42.70) & 426 (39)\\\hline
10\% & 1 (1) & 426 (214) & 222.17 (91.69) & 166 (84) \\\hline
20\% & 3 (3) & 426 (409) & 242.14 (183.55) & 223 (171) \\\hline
60\% & 12 (12) & 1207 (1207) & 532.90 (544.20) & 449 (493) \\\hline
80\% & 17 (12) & 1594 (1594) & 704.31 (726.30) & 609 (656) \\\hline
\end{tabular}
\vspace*{-.2\baselineskip} \caption{\small Node Degree in the presence of RSUs
(in parenthesis w/o RSU) under different penetration ratios.}
\label{table:RSUDegree}
\vspace{-7mm}
\end{table}

Next, we examine the evolution of the VANET graph \emph{\textbf{effective
diameter}} over time. Primarily, we find that the effective diameter of both the
real and realistic scenarios follows the diameter of the biggest cluster in the
network. For the Los Angeles scenario, the effective diameter exhibits high
variance for $~\emph{P}<10\%$ without exceeding 9 hops. Exceeding~\emph{P}=10\%,
the effective diameter stabilizes to around 5 hops. For the Shanghai scenario,
when $~\emph{P}<40\%$, the effective diameter is also low and quite stable(~ 2
hops). This is due to poor network connectivity caused by the presence of
several small clusters. On the other hand, as the penetration ratio increases
over~\emph{P}=40\%, the effective diameter increases and stabilizes to around 9
hops.

Continuing our network analysis, we find that the VANET~\emph{exhibits small
world properties}. This arises from the observation that the
\emph{\textbf{average degree of separation}} (i.e., mean shortest path length
between pairs of nodes) over time, remains low for all the scenarios we studied.
For instance with~$\emph{P}=100\%$, the average degree of separation between two
vehicles is 3.3 hops for the Los Angeles area and 5.2 hops for the Shanghai area
with smooth changes as time progresses and with very small variability. The
presence of RSUs has no effect on the average degree of separation,
when~$\emph{P}>10\%$. Moreover, in both scenarios our analysis has indicated
that the average degree of separation increases linearly with the average
geographic distance of node pairs but never exceeds the above values.

Table~\ref{table:Densities} presents the VANET connectivity under different
penetration ratios for the Los Angeles and Shanghai areas. For the Los Angeles
scenario, the effective diameter and the average degree of separation do not
present significant changes. This is due to the fact that the number of vehicles
(even when~\emph{P}=20\%) is high enough to allow the VANET fusion, that is,
many clusters which are present when ~$~\emph{P}<20\%$, are now fused into
larger clusters, thereby enabling even more vehicles to communicate among them.
For the Shanghai scenario, this fusion process is evident when ~$\emph{P}>20\%$,
and consequently, the effective diameter and average degree of separation do not
present significant changes for $\emph{P} \ge 80\%$ (not shown here).


\begin{table}[t] \tiny \centering \begin{tabular}{|l|c|c|c|c|c|}\hline
\multicolumn{6}{|c|}{\bf Los Angeles (Realistic)} \\\hline Penetration & {\bf
10\%} & {\bf 15\%} & {\bf 20\%} & {\bf 40\%} & {\bf 60\%}\\\hline\hline
Effective Diameter & 5.6 & 5.5 & 5.3 & 5.1 & 5.0 \\\hline Node Degree (Median) &
84.0 & 123.0 & 171.0 & 325.0 & 493.0  \\\hline Avg. Degree of Separation & 3.5 &
3.4 & 3.4 & 3.3 & 3.3  \\\hline\hline \multicolumn{6}{|c|}{\bf Shanghai (Real)}
\\\hline Effective Diameter & 1.0 & 1.7 & 4.3 & 5.1 & 8.9 \\\hline Node Degree
(Median) & 1.5 & 1.5 & 2 & 4 & 6.0 \\\hline Avg. Degree of Separation & 1.0 &
1.38 & 2.6 & 5.4 & 5.8 \\\hline
\end{tabular}
\vspace*{-.2\baselineskip} \caption{\small Effective Diameter, Node Degree and
Avg. Degree of Separation under different penetration ratios.}
\label{table:Densities}
\vspace{-7mm}
\end{table}

\textbf{Lessons:} 1) The degree distribution of VANETs is positive skewed at
low value of penetration ratio. With the increase of penetration
ratio, this skewness fades as the majority of nodes have high degree and network
connectivity improves. 2) The VANET communication graphs exhibit small world
properties. This indicates that most node pairs are connected by at least one
short path, indicative by the average degree of separation which remains low,
independent of penetration ratio. In addition, there are several vehicles (hubs)
in the network with a very large degree, that mediate large path lengths between
other node pairs.

These observations are useful since they provide indications about the richness
of the network connectivity. Gossip protocols~\cite{Boyd2006} should be
investigated in order to take advantage of small-world properties, and
especially scale-free ones, in order to efficiently disseminate information in
vehicular networks. In addition, the network's density makes the use of power
transmission adjustment mandatory. The difference in a VANET setting is that
this procedure must be {\it continuous}, and can not be decided once in advance.
Such continuous power adjustment is not easy to achieve, unless it is done on a
per-road-segment basis (e.g., CVIS platform).


\subsection{Cluster Analysis}
\label{subsec:ClusterAnalysis}

Network analysis performed in the previous subsection provided evidence that
vehicles are partitioned into several clusters. Therefore, it is imperative to
perform a thorough study of these clusters and understand their properties as a
function of time and transmission range. Later in this subsection, we look
deeper into the VANET communication graph to examine the existence of
communities (dense sub-graphs) and their dynamics.

As seen on Subsection~\ref{subsec:NetworkAnalysis} the penetration ratio
influences significantly the \emph{\textbf{number of clusters}} existing in the
VANET. Specifically, for the real scenario, at~\emph{P}=20\%, the number of
clusters is \textit{approximately 5 times larger} than that for~\emph{P}=100\%.
Moreover, in the former case the biggest cluster (in terms of membership)
contains at most only 30\% of the total vehicles.
In such cases the placement of static RSUs can efficiently bridge fragmented
parts of the VANET, reducing thereby the number of clusters and improving the
overall end-to-end connectivity. For this scenario only, we simulated the
placement of 427 RSUs in the area of Shanghai, following the same density per
$km^2$ as found in the Los Angeles area. The results indicated a reduction of up
to 22\% in the number of clusters present in the VANET is possible. On the other
hand, when $\emph{P}>60\%$, the number of clusters remains constantly less than
7 (with low variability) with the size of each cluster being significantly
large. The case of the realistic scenario however exhibits different behaviour
from the above, with penetration ratio changes having minimal influence to the
number of clusters present in the network. This can be accounted to the fact
that vehicular density is significantly high, and a transmission range of 300m,
even when $\emph{P}<20\%$, is enough to form 1 single large cluster, containing
the majority of vehicles.

The \emph{\textbf{average clustering co-efficient}} for the real scenario
fluctuates around~68\% for~\emph{P}=100\% and gets smaller and notably variant
as vehicle penetration decreases. In order to compensate to these side-effects
and keep a relatively stable clustering co-efficient close to the above value,
our study has indicated that a minimum density of 5 RSUs per $Km^2$ is required.
Concerning the realistic scenario the clustering co-efficient stabilizes to
about 71\% for all values of vehicle penetration. It is worth noting that,
despite the difference in vehicle density among the two scenarios, the average
clustering co-efficient remains relatively close and consistent. This denotes
that once a specific level of vehicle density is reached in the VANET, the
internal connectivity of the formed clusters is not affected by changes in the
vehicle mobility pattern due to different underlying road topology.
Nevertheless, even in low vehicle densities, the clustering co-efficient is a
good indicator in terms of the connectivity of the nodes contained in the
cluster. Hence, nodes that belong to a cluster with a higher co-efficient than
others, will be prefered for the forwarding of information in sparse networks,
since there is a higher propability of establishing a communication path.


\begin{figure}[t]
    \centering
     \subfigure []{
        \includegraphics[width=0.32\textwidth]{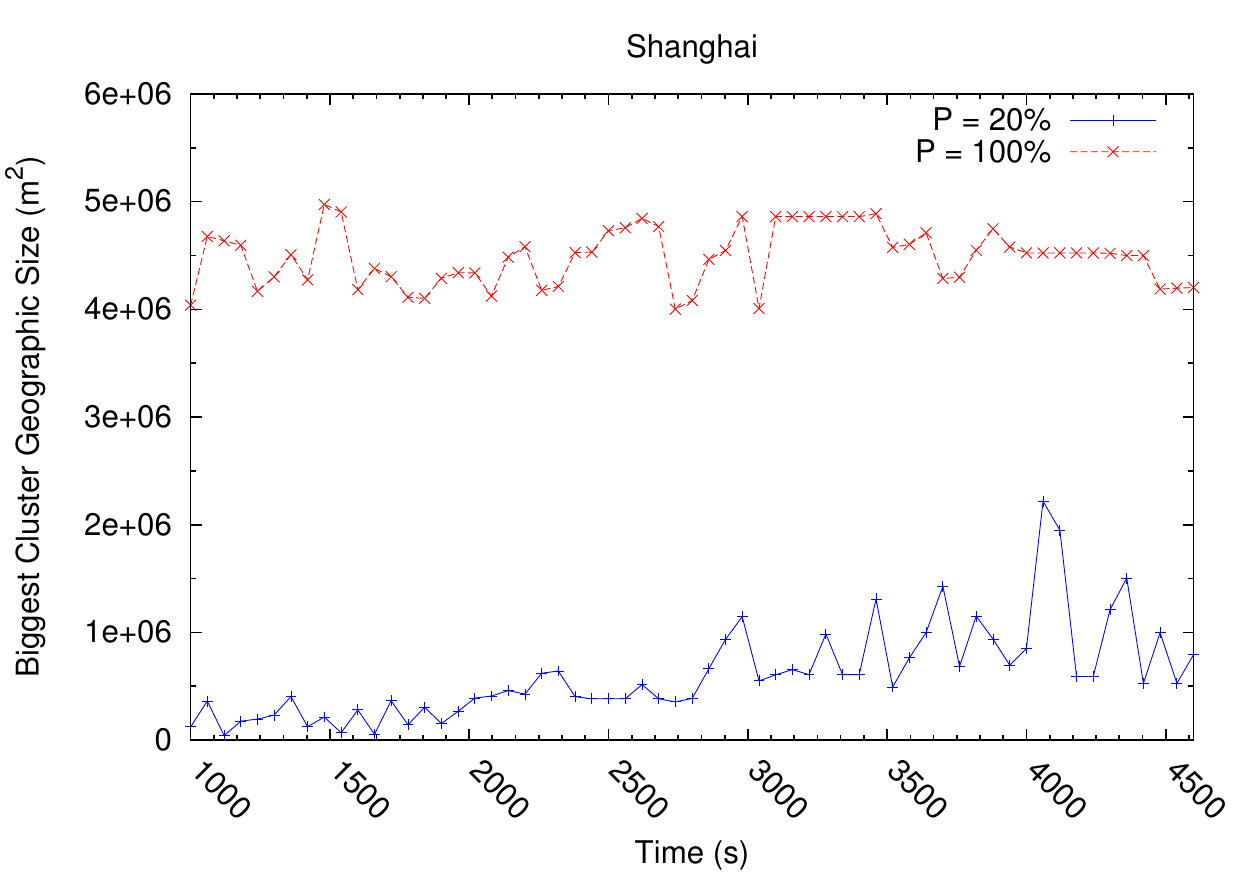}
        \label{fig:bigClustGeoSizeReal}
   }
    \subfigure []{
        \includegraphics[width=0.32\textwidth]{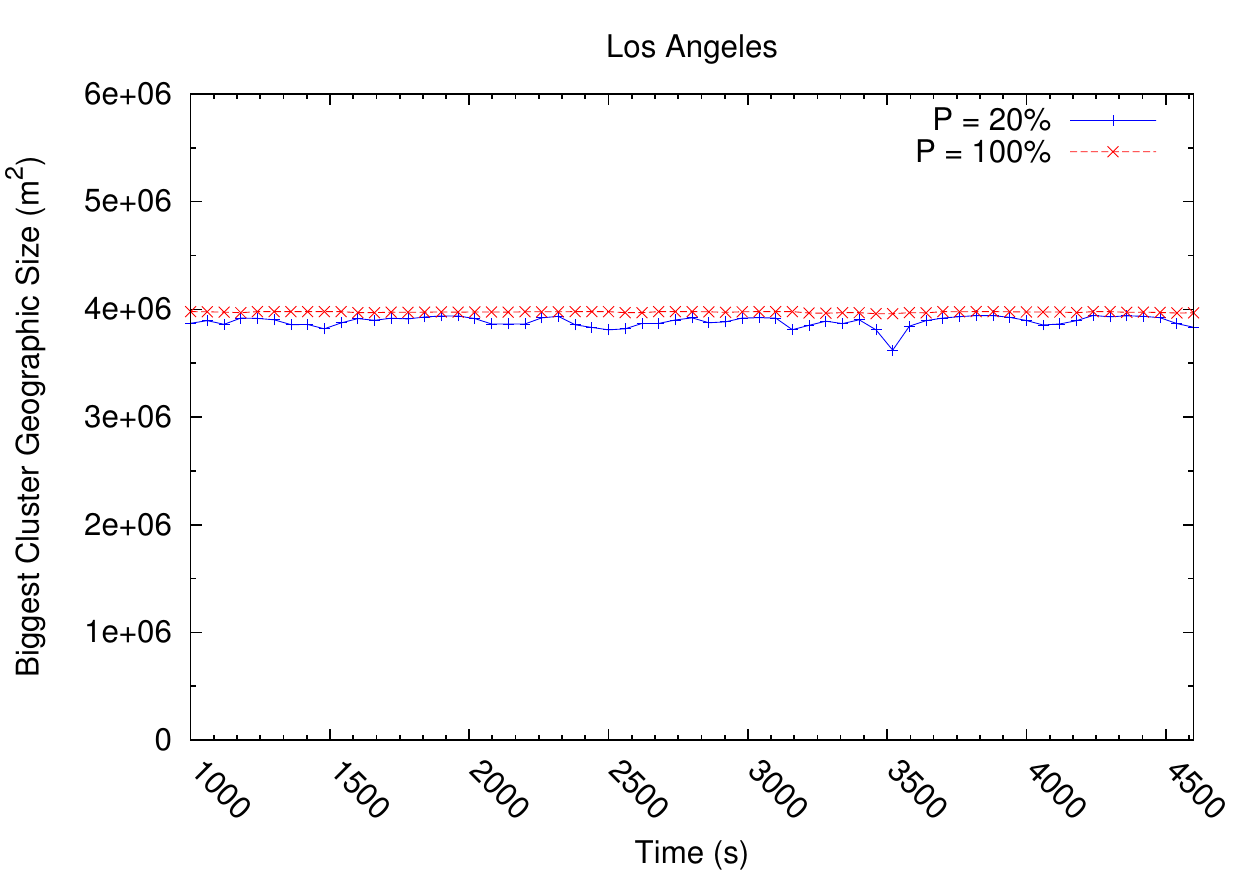}
        \label{fig:bigClustGeoSize}
    } \caption[]{Biggest Cluster Geographic Size vs. Time.}
    \vspace{-6mm}
\end{figure}


As previously mentioned, when $\emph{P}>60\%$ the VANET graph is comprised of
few clusters, mainly one large cluster containing the majority of vehicles and
other significantly smaller clusters in its surroundings. To further understand
the \emph{\textbf{biggest cluster}} in terms of the number of vehicles, we
choose to study its properties as a function of time and penetration ratio. To
this end, Figures~\ref{fig:bigClustGeoSize} and ~\ref{fig:bigClustGeoSizeReal}
present the biggest cluster geographic size in the areas of Los Angeles and
Shanghai respectively, at the extremities (\emph{P}=20\% \& \emph{P}=100\%) of
vehicle penetration. Initial observations indicate that the geographic size of
this cluster exhibits similar behaviour in both areas, thereby concluding that
the road topology has little or no effect on its structure and evolution. In
particular, for~\emph{P}=20\% the biggest cluster occupies almost the whole Los
Angeles area and approximately half of the whole Shanghai area. In low-vehicle
density environments such as the one in Shanghai, the biggest cluster in terms
of geography, tends to be small in population size, thereby being prone to
changes (visible in the lower part of Figures~\ref{fig:bigClustGeoSizeReal}) due
to the arrival and departure of vehicles. Interestingly, the utilization of RSUs
allows the biggest cluster in the Shanghai area to expand massively in size and
occupy $\geq87\%$ of the space (not shown in the aforementioned figure). On the
other hand, the geographic size of the biggest cluster when~\emph{P}=100\%, in
the area of Los Angeles, presents a smooth and less variant change in time,
while occupying almost 99\% of the given area. These levels of coverage are
significant, since vehicles participating in the largest cluster, can
disseminate information in large and remote geographic areas. In terms of
cluster membership, in the Shanghai area, for~\emph{P}=20\%, the biggest cluster
contains between 10\% to 33\% of the total vehicle, while for the realistic
scenario it contains more than 98\% of the total vehicles.

However, despite the usefulness of the aforementioned findings, the geographic
coverage of the biggest cluster alone does not accurately capture its quality in
terms of providing end-to-end connectivity and effective information
dissemination. To complement the above information and better quantify the
quality of the biggest cluster, we investigate its \emph{\textbf{local
clustering co-efficient}} as a function of time and market penetration ratio. We
discover that at low vehicle densities, the biggest cluster co-efficient is
quite variant and at several instances, is significantly lower than the average
clustering co-efficient. As the penetration ratio increases the local
co-efficient remains stable for both scenarios close to 70\%. This observation
is important since it indicates the high density of communication links in the
cluster and subsequently the existence of almost-cliques in the network, despite
the different undrelying road topology.

We proceed our analysis by exploring the VANET communication graph for the
existence of \emph{\textbf{communities}}, that is groups of vehicles with more
intra-communication links than inter-communication links. Given the high density
of intra-communication links, information replication could easily be achieved,
allowing communities to act as ``data islands''.

Community analysis begins by investigating the modularity of the VANET
communication graphs. \emph{\textbf{Community modularity}} quantifies in the
range of -1 to 1 the division of the graph into communities, with good divisions
giving communities with dense internal connections and weak connections between
different communities. We use the Girvan-Newman algorithm~\cite{Clauset2004} in
order to identify the communities in VANET. For the real scenario, and
at~\emph{P}=20\%, community modularity is quite variable and approximates the
value of 0.8, indicating very good modularity. Despite the fact that in this
case, the VANET exhibits excellent modularity in such conditions, this is
misleading since low penetration ratios highly partition the VANET with the
majority of vehicles belonging to different partitions and connected only to a
handful of other vehicles. One could argue that such small communities could
still be used to ferry information, but our study has shown that the highly
dynamic nature of the VANET causes them to dissolve rapidly. On the other hand,
as, for~\emph{P}=100\% community modularity does not fall below 0.63.

\begin{figure}[t]
 \centering \includegraphics[width=3.2in]{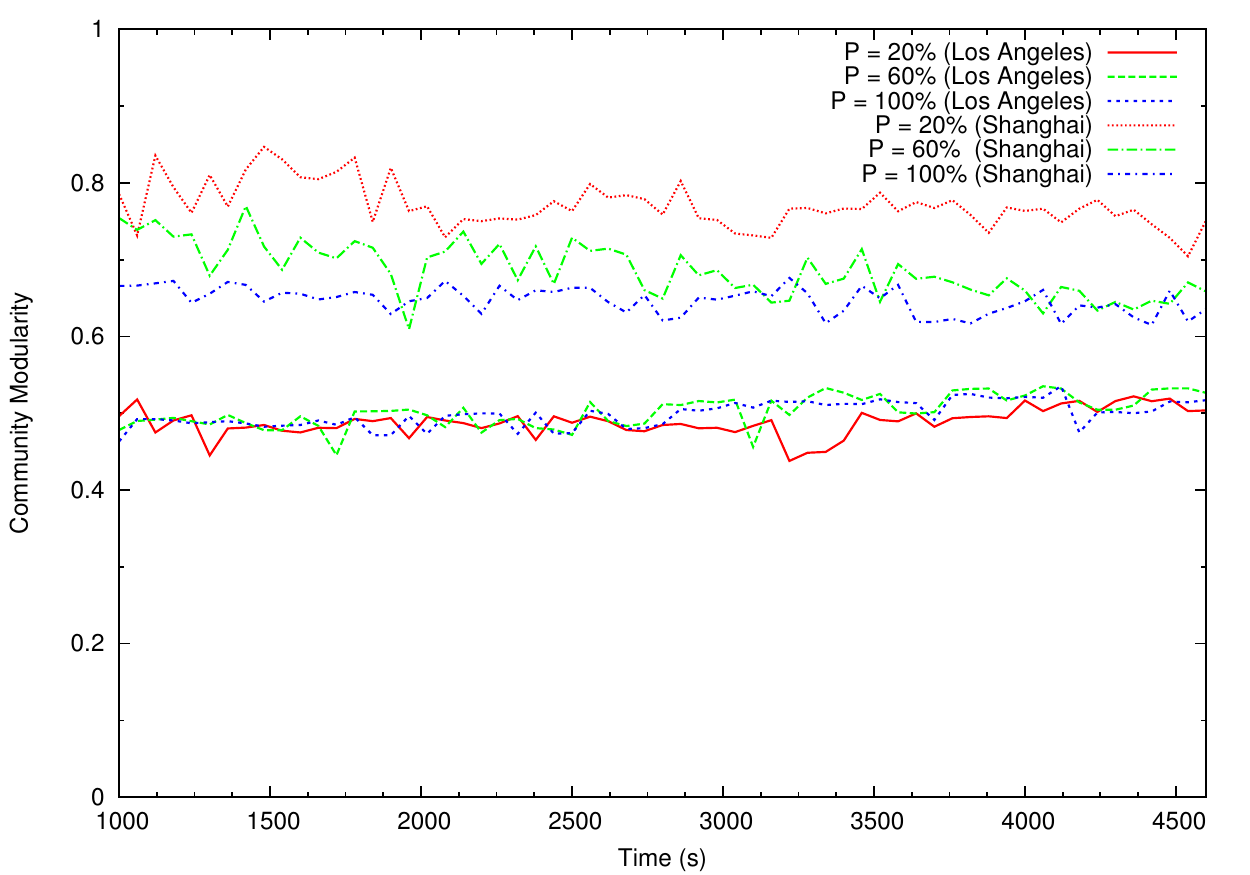}
\caption{Community Modularity, under different penetration ratio.}
\label{fig:CommModularity}
\vspace{-6mm}
\end{figure}

As Figure~\ref{fig:CommModularity} indicates, for the Los Angeles scenario
community modularity stabilizes to approximately 0.48 and does not fall below
0.45, irrespective of the penetration ratio. This shows that even with lower
modularity values, a good division of the VANET can be achieved and identify
effectively the candidate ``data islands''. Therefore, information dissemination
protocols could ultimately use such structures to ferry information from one
location to another with low probability of delivery failure.


\textbf{Lessons:} 1) At low penetration ratios and low vehicle densities, even
a small number of RSUs per $Km^2$ are sufficient to maintain stable clustering
coefficients over time. Maintaining a stable clustering coefficient allows node
pairs in a cluster to maintain connectivity. 2) Strong cluster connectivity has
the potential of improving end-to-end network connectivity in the presence of
vehicles that bridge one or more clusters. At high penetration ratio, the
clustering coefficient is very high and stable, and the introduction of RSUs has
minimal benefits. 3) The clustering co-efficient provides a good indicator about
the connectivity of the nodes contained in the associated cluster, even in low
vehicle density. As we will show in the next section, the clustering
co-efficient can serve as a criterion for selecting nodes that can forward
information, at low vehicle density environments. 4) At low vehicle densities,
the biggest cluster in terms of membership, does not correspond to better
network connectivity. Consequently, nodes that belong to the biggest cluster may
be less preferable for forwarding information. 4) VANET includes small
communities which can be combined into larger sets of nodes that can also be
meaningfully interpreted as communities.

The existence of communities implies that mobility models like the Random Way
Point, which are based on types of random walks should be abandoned, because
they do not produce clusterings of the vehicles and additionally they do not
support the existence of ``hub" vehicles that explain the distributions of the
centrality metrics. Therefore, research towards richer models (e.g.,
\cite{Musolesi2007}) must be conducted. Specifically, vehicles that belong in a
community can be seen as a ``data island''. This is widely observed in highways,
where opportunistic protocols are used (e.g., Opportunistic Packet Relaying
protocol - OPERA~\cite{Abuelela2008}). Therefore, since we also observe
communities in urban environments, such protocols could also be adapted to urban
environments. Furthermore, the existence of communities implies also that leader
election algorithms will work successfully~\cite{Raya2006}, especially if we
incorporate the centrality metrics in their selection. In addition, the
existence of communities contributes on addressing the Maximum Coverage
Problem~\cite{Trullols2010}. Considering an area with an arbitrary road topology
that must be equipped with a limited number k of RSUs, where to place them so as
to maximize the dissemination of information? In points where borders of
clusters exist, or in places where vehicle communities exist. Similarly,
installation of RSUs is suggested in places where the nodes have low localized
clustering coefficient (sparse network), and thus the delivery of messages would
require a significant amount of time without the infrastructure.

\subsection{Centrality Analysis}
\label{subsec:Centrality}

We proceed with our analysis by investigating the existence and spatial
distribution of ''central'' nodes in the VANET communication graph. An initial
observation for all the scenarios under study is the following: \textit{low
penetration ratios do not foster the existence of ''central'' nodes in the
VANET}. Recall, that the definition of \emph{\textbf{lobby index}} implies that
it is a generalization of node degree and some sort of ''simplification'' of
betweenness centrality. Therefore, this behaviour is expected, since as per our
findings in the previous sections, in such situations the network is sparse and
lacking the existence of large clusters, which in turn means that the majority
of vehicles are connected only to very few other vehicles in their close
proximity. However, this is not the case for the realistic scenario. Since
vehicle density is significantly higher than in the respective scenario of
Shanghai, vehicles have a considerable high value of lobby index with relatively
variant in the course of time. The presence of RSUs, however, allows vehicle
centrality to stabilize in the aforementioned case.  Due to their static
position, RSUs function as relay points through which communication paths
between pairs of vehicles are established. Effectively, all RSUs have the same
and significantly higher lobby index (around 240) from all other nodes in the
VANET. Furthermore, our analysis has indicated that in scenarios with low
penetration ratio,  \emph{the vehicles that exhibit high lobby index than the
rest vehicles, are the ones which are at the vicinity of a road intersection}.



\begin{figure*}[ht]
    \centering
     \subfigure[Shanghai P=20\%]{
     \includegraphics[width=0.30\textwidth]{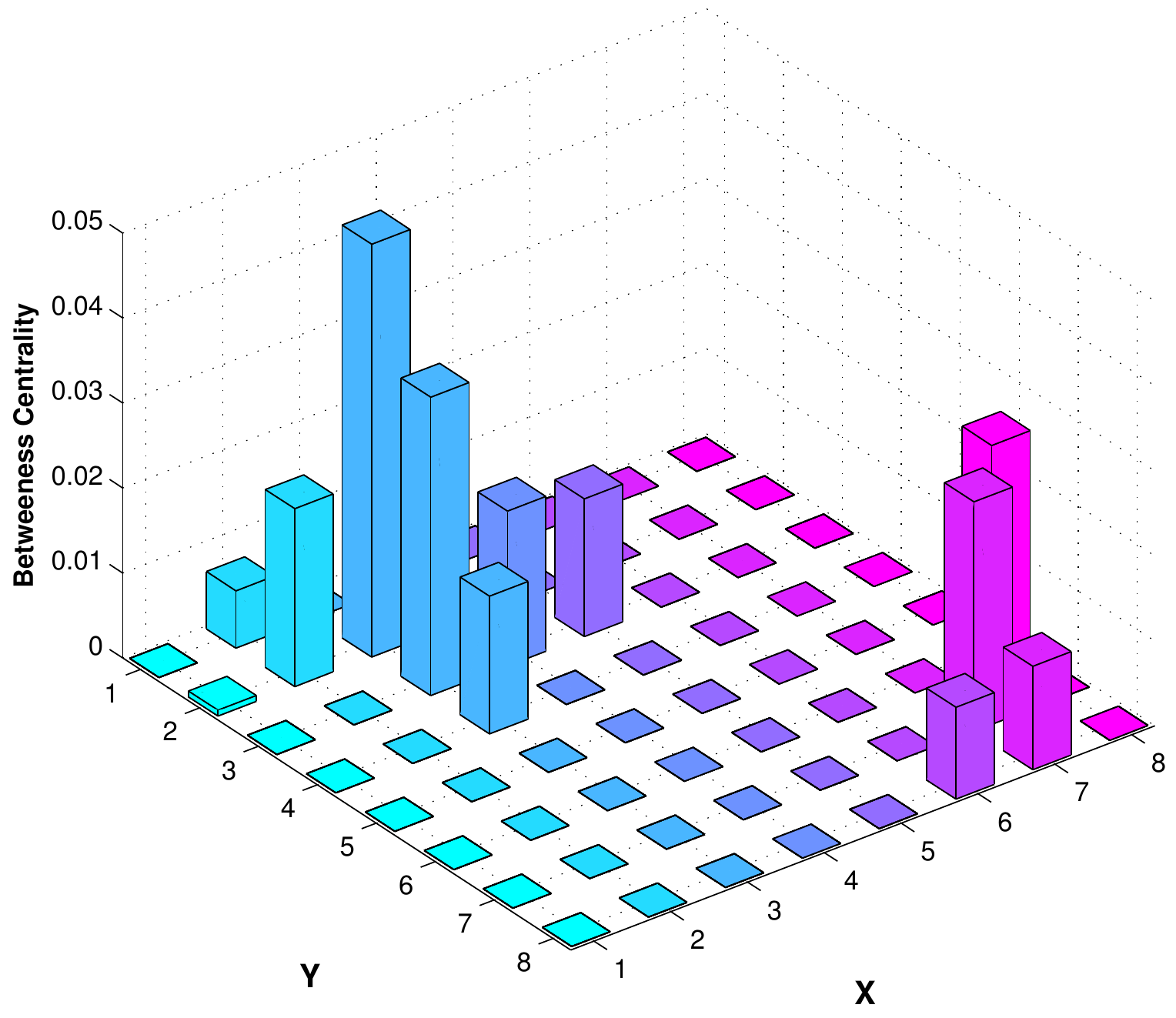}
    }
     \subfigure[Shanghai P=60\%]{
     \includegraphics[width=0.30\textwidth]{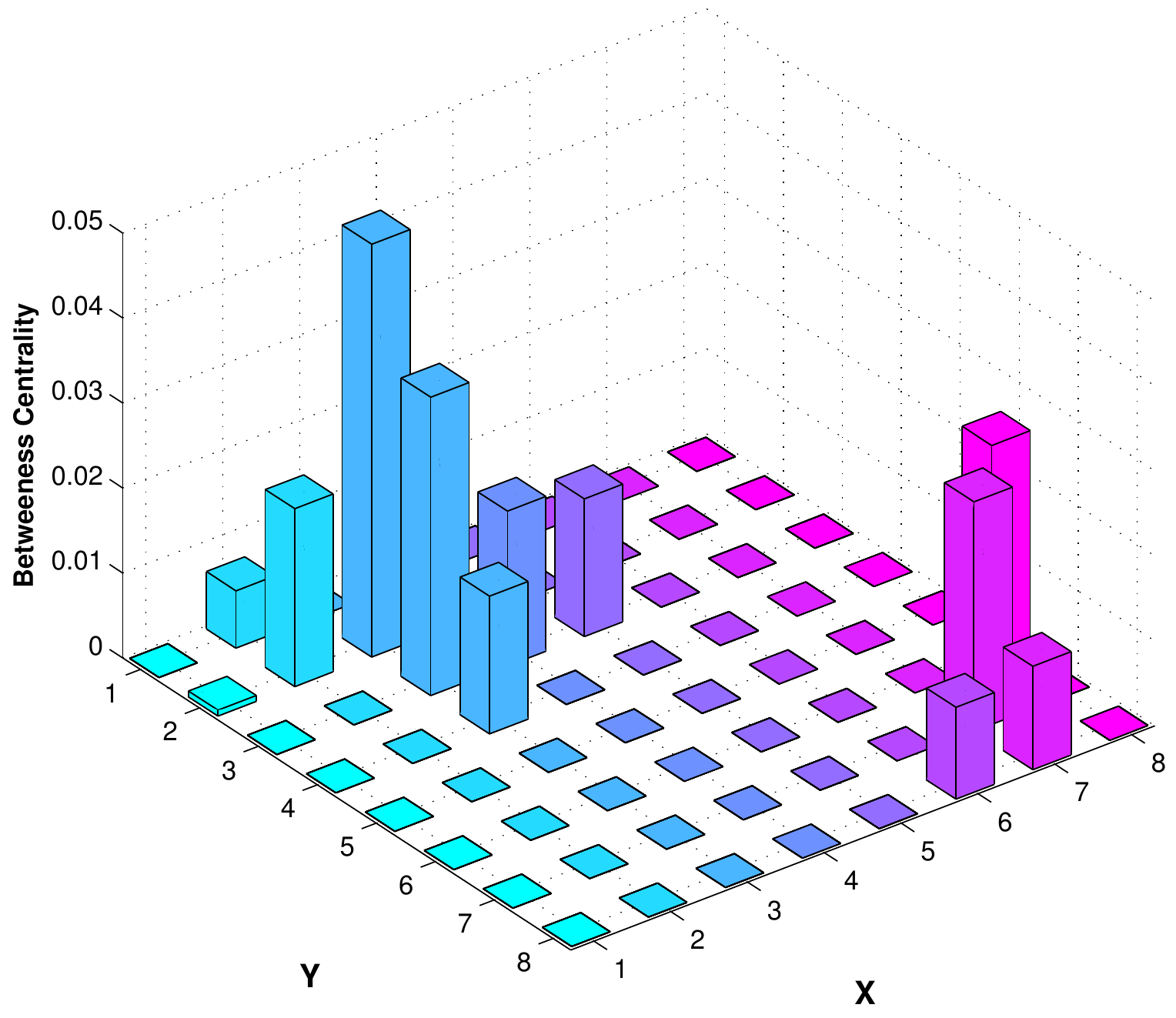}
    }
     \subfigure[Shanghai P=100\%]{
     \includegraphics[width=0.30\textwidth]{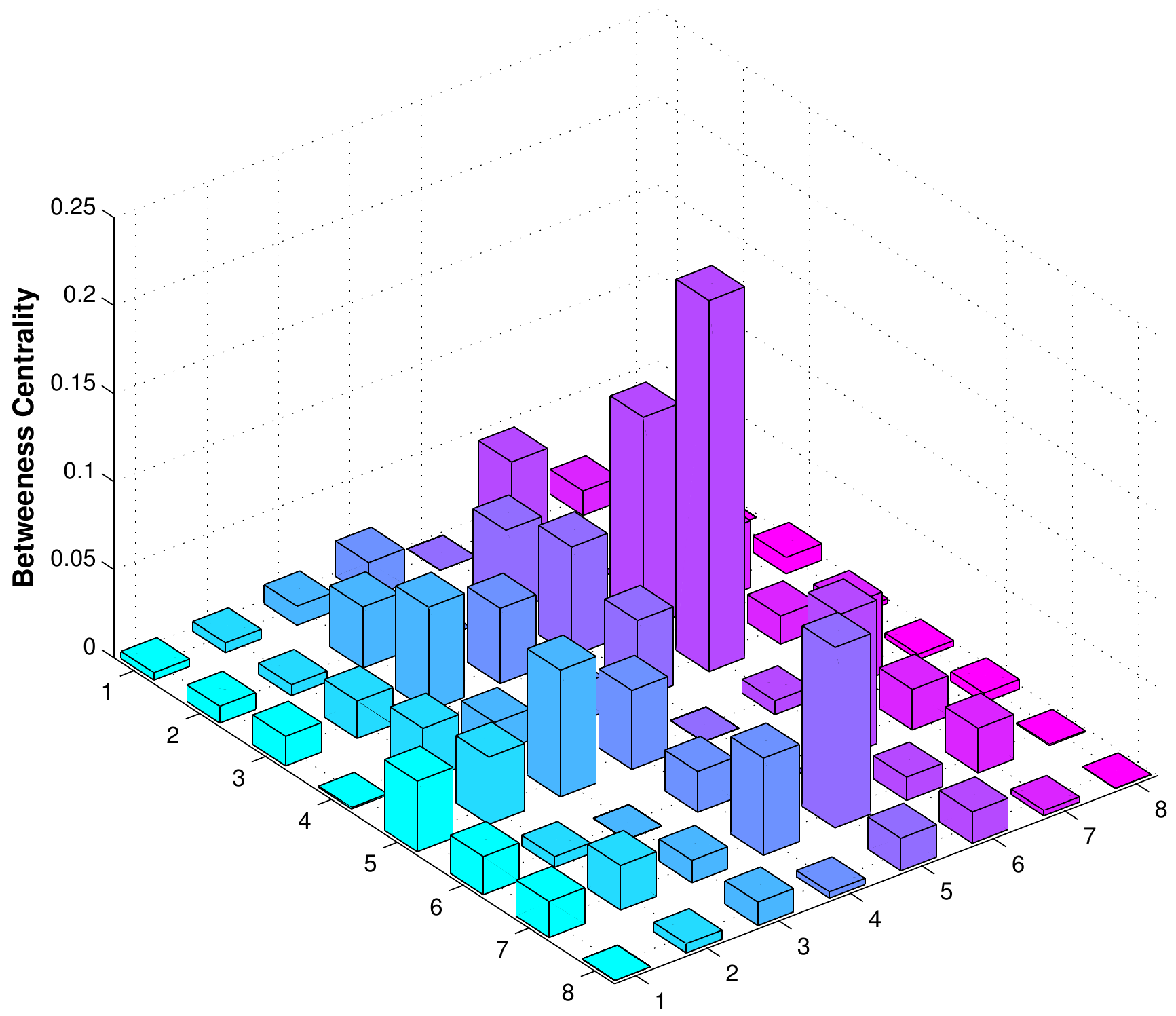}
    }
    \\
     \subfigure[Los Angeles P=20\%]{
     \includegraphics[width=0.30\textwidth]{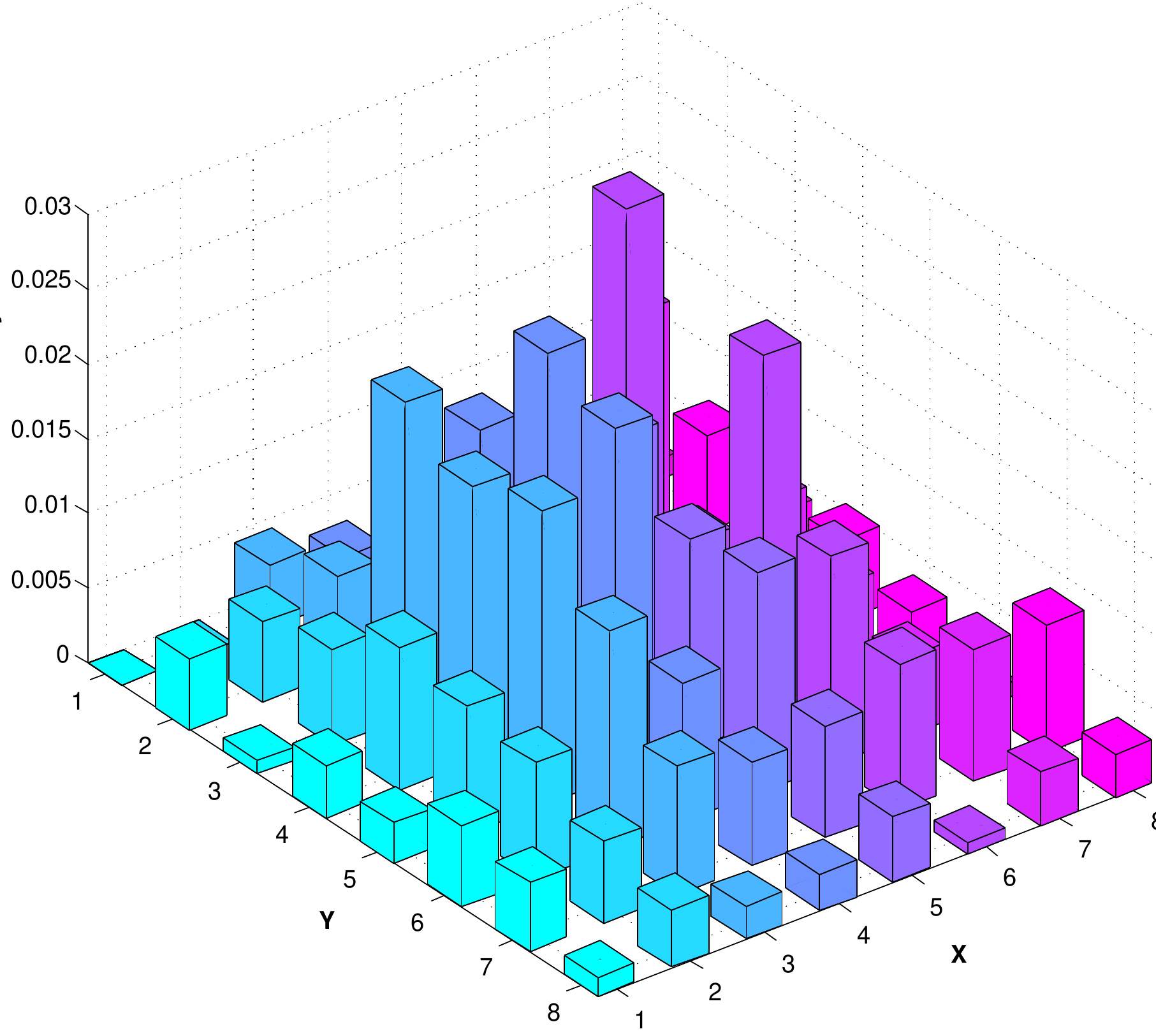}
    }
     \subfigure[Los Angeles P=60\%]{
     \includegraphics[width=0.30\textwidth]{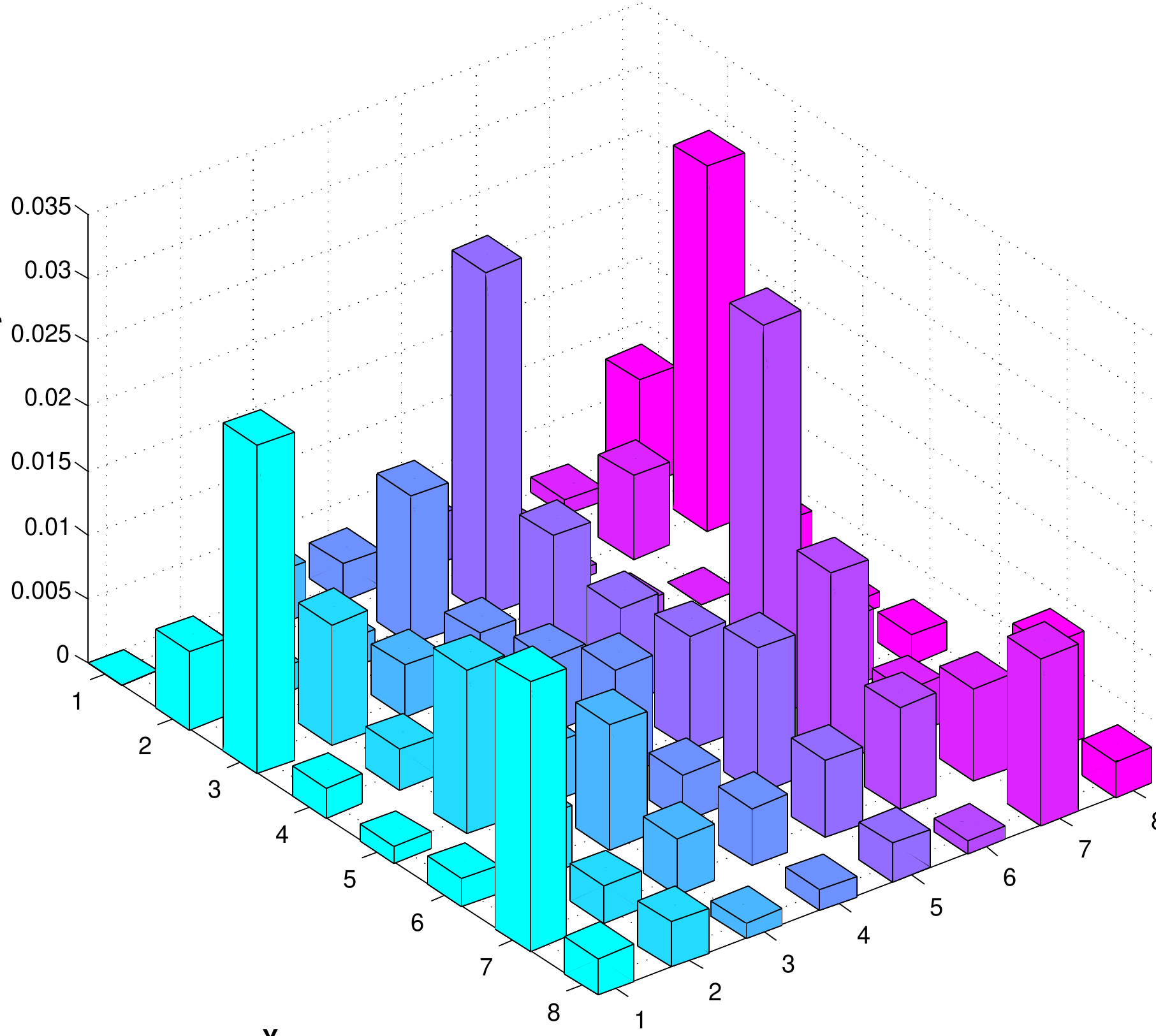}
    }
     \subfigure[Los Angeles P=100\%]{
     \includegraphics[width=0.30\textwidth]{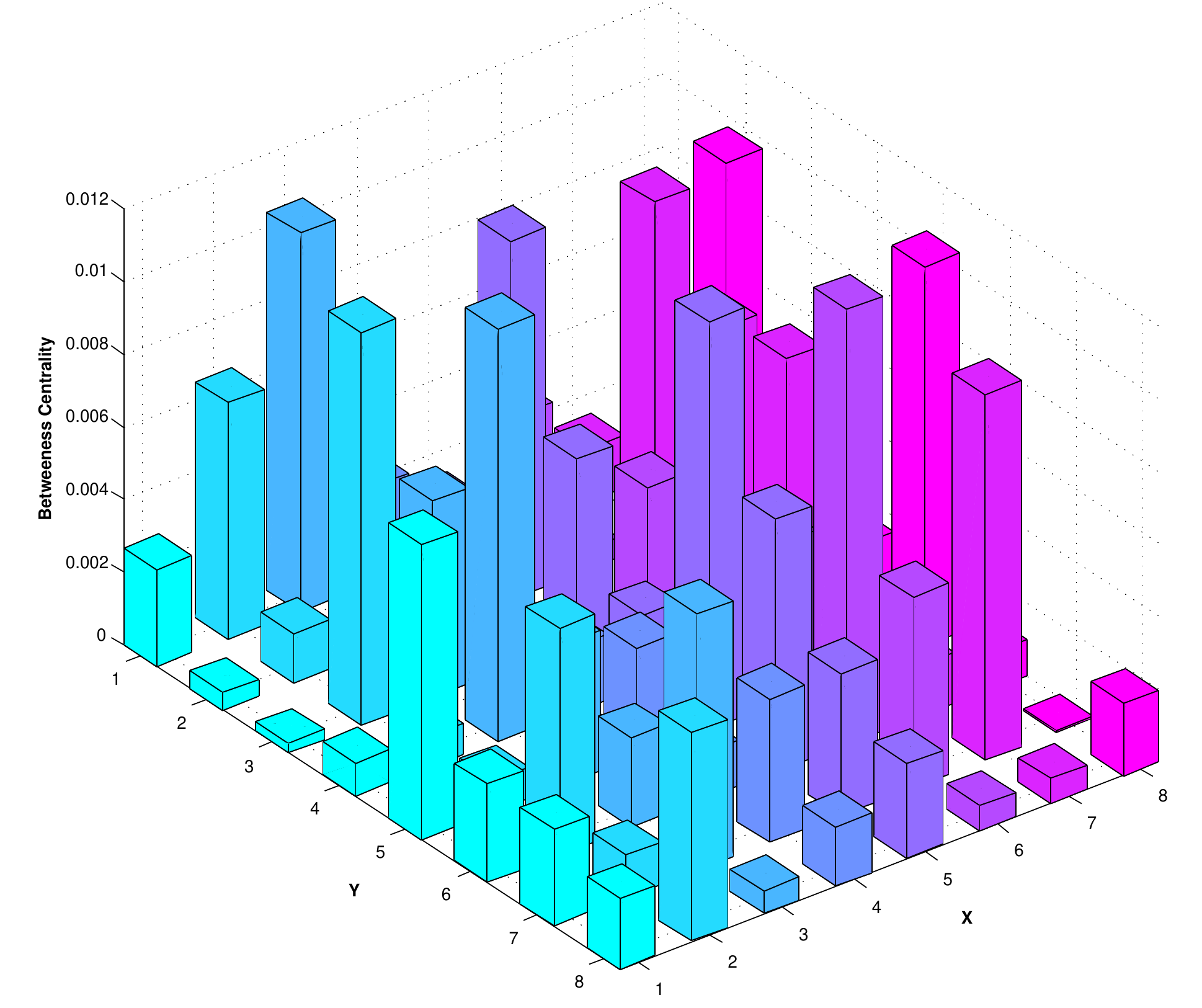}
    }
    \caption{Geographic Distribution of Betweenness Centrality for Los Angeles
    and Shanghai.}
    \label{fig:geoBC}
\vspace{-4mm}
\end{figure*}

Figure~\ref{fig:geoBC} provides a snapshot of the Los Angeles and Shanghai areas
divided into 64 rectangular zones, with each zone covering a geographic area
equal to $250 m2$. For each zone, the \emph{\textbf{betweenness centrality}} of
the contained nodes is normalized, averaged and plotted on the z-axis. As
Figure~\ref{fig:geoBC}(a) depicts, at~~\emph{P}=20\% the majority of zones have
average betweenness centrality very close or equal to 0.025. The zones which
exhibit high levels of betweenness centrality are the ones containing densely
populated intersections, which in turn allow vehicles to establish more
connections to other vehicles from what would be the case in other locations.
Unexpectedly, even with the much higher vehicle density of the Los Angeles
scenario, the overall betweenness centrality of the nodes in the network is
slightly increased, however, the geographic distribution is extended
considerably.


\textbf{Lessons:} 1) Lobby centrality index is a better metric for identifying
the ''central'' nodes in the VANET than the betweenness centrality metric. This
is due to the statistic nature of the betweenness centrality metric, since most
nodes have low and same betweenness centrality values. 2) Low penetration ratios
do not foster the existence of ''central'' nodes in the VANET. The high mobility
of vehicles, make graph-based topology-control methods (spanning trees,
Ga\-briel/Yao graph, RNG) not appealing, whereas, clustering is preferable.
Which could be the clusterheads? Not necessarily high-degree nodes, but those
with large betweenness centrality (if we need a few clusters), or those with
high lobby centrality (if we need a lot of clusters). 3) The road network alone
is not sufficient information to identify the positions of possible
``significant'' nodes in a VANET.

\subsection{Link Analysis}
\label{subsec:LinkAnalysis}

The \emph{\textbf{link level analysis}} of the VANET communication graph
contributes to the prediction of the network-link lifetime. The number and
duration of connected periods between any two vehicles, as well as the duration
between successive connected periods is influenced by driving situations and
vehicle speed. Table~\ref{tbl:linkDuration} present the link-level statistics
for the Los Angeles and Shanghai areas when ~\emph{P}=60\%. \begin{table}[t]
\tiny
\begin{center}
\begin{tabular}{|l|c|c|c|c|}\hline
\multicolumn{5}{|c|}{\bf Shanghai} \\\hline
{} & {\bf Min}   & {\bf Max} & {\bf Mean} & {\bf Median} \\\hline\hline
Number of Connected Periods		& 1 & 11 & 1.7 & 1 \\\hline
Link Duration	& 1 sec & 4541 sec & 108.27 sec & 41 sec \\\hline
Re-Healing Period				& 1 sec & 4550 sec & 370.16 sec & 47 sec \\\hline\hline
\multicolumn{5}{|c|}{\bf Los Angeles} \\\hline
Number of Connected Periods		& 1 & 43 & 2.39 & 2 \\\hline
Link Duration	& 1 sec & 4503 sec & 393.44 sec & 226 sec \\\hline
Re-Healing Period				& 1 sec & 4500 sec & 979.85 sec & 463 sec \\\hline
\end{tabular}
\end{center}
\caption{\small Link level statistics for Shanghai and Los Angeles scenarios at
P=60\%.}
\label{tbl:linkDuration}
\vspace{-6mm}
\end{table}

It is evident that higher vehicle densities, as it is the case of the realistic
Los Angeles scenario, increase the time period in which two vehicles are in
range of each other, thus allowing established links to have a longer duration.
An increased link duration reduces the number of connecting periods between the
two vehicles and consequently minimizes any overhead imposed by the process of
re-establishing connections. Specifically, we observe that the duration of the
connected periods in the Los Angeles scenario is almost 3x as the one in the
Shanghai scenario. In general, results show the connectivity between two
vehicles changes very frequently. In addition, it is worth noting that by
comparing the mean and median, we can conclude that  \emph{there is a
significant high variability in the link duration values for both scenarios}.

\begin{figure}[t]
 \centering
 \includegraphics[width=3.2in]{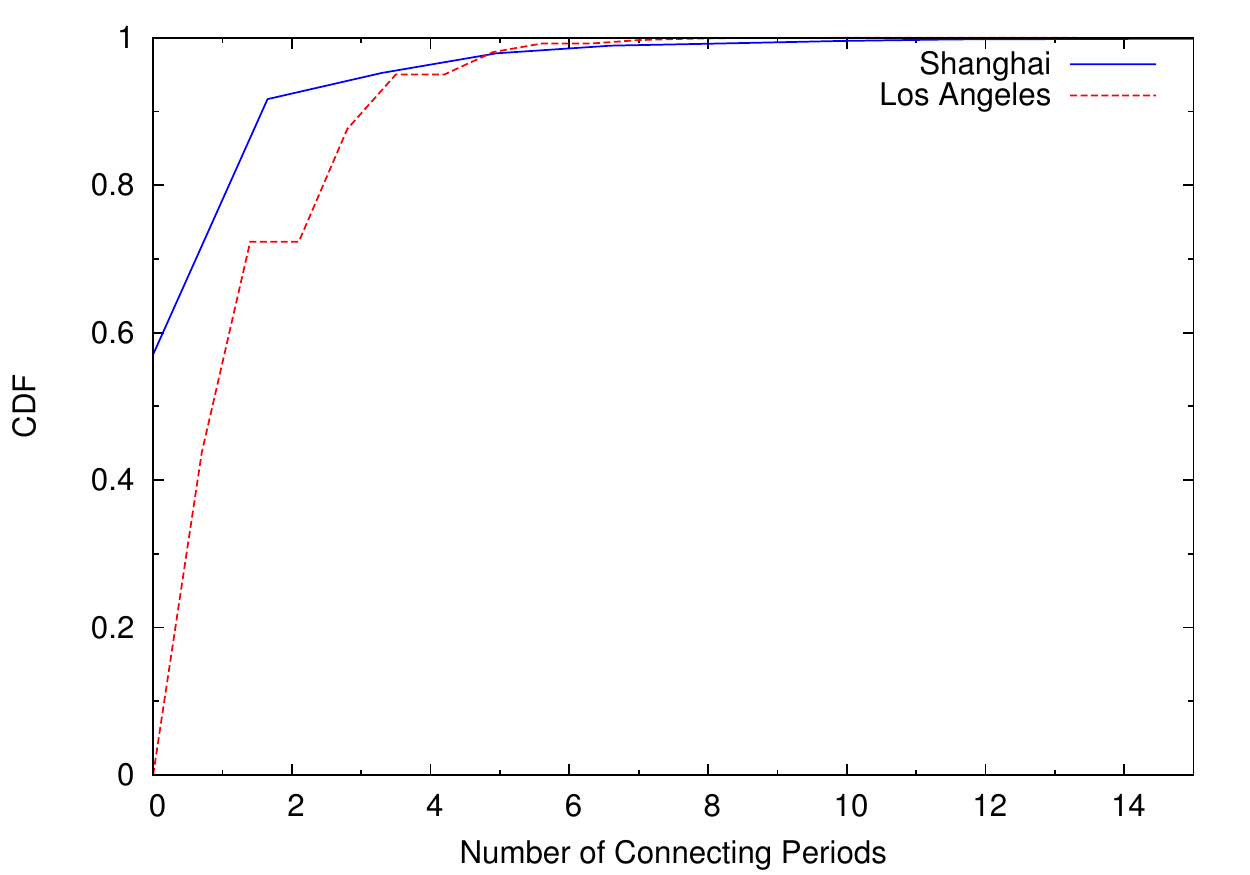}
 \caption{Connecting Periods CDF at P=60\%}
 \label{fig:ConnectingPeriods}
 \vspace{-6mm}
\end{figure}

Figure~\ref{fig:ConnectingPeriods} indicates that the number of connecting
periods for both the real and the realistic scenarios remains relatively low for
most vehicles. In particular, we note that there is a very low probability the
same two vehicles will meet in excess of 3 times. 

\textbf{Lessons:} 1) The network connectivity is highly dynamic and it is
affected by the network density. Specifically, the link duration of vehicles is
almost tripled at high vehicle densities, whereas, the connected periods is
almost doubled at high vehicle densities. This implies that a limited number of
''killer applications'', (i.e., large data files transfer and real-time
information dissemination) can be supported by VANETs. Nonetheless, as we will
show in the next section, one way to increase the set of applications that can
be supported is to enhance VANET routing protocols with VANET graph knowledge.
2) Most vehicles have low number of connected periods. This means that it is not
required to re-initiate often the route discovery process in routing protocols.
Also, the duration of connected periods is important for designing a routing
protocol, since several applications in VANETs are governed by the duration of
the connected periods and how frequently the path between vehicles becomes
unavailable. Specifically, it can be utilized to provide information exchange
resilience by estimating an upper boundary to the amount of information that can
be exchanged between any given pair of vehicles. 3) The re-healing time between
vehicles is large for both the real and the realistic scenarios in urban
environments; this observation is useful since it can be used to indicate the
size of message buffer and how long a vehicle should buffer the broadcast
message before it rebroadcasts it to another vehicle. Also, this metric is
useful for determining how frequently a vehicle should periodically send out the
beacon messages so that the relay vehicles can be discovered without flooding
the network. Finally, the re-healing time measure is a critical metric for
implementing carry-and-forward protocols for VANETs. In these protocols,
vehicles buffer the broadcast message for some period before broadcast it to
other vehicles.
\section{Implications}
\label{sec:simulation}

In the previous section we conducted a thorough analysis of the spatio-temporal
characteristics of a large-scale VANET graph and gained a deep understanding of
its shape. The question that remains to be answered is whether this knowledge
is indeed useful from an engineering perspective. In this section, we seek to
answer this question by providing two known VANET routing protocols,
VADD~\cite{Zhao2006} and GPCR~\cite{Lochert2005} with VANET graph information,
to determine whether such information can improve their quality of service. We
opted to use the above routing protocols, since each one is a member of a
different family of VANET protocols with its own set of
requirements~\cite{Lin10}.

\subsection{VADD: Vehicle Assisted Data Delivery}
VADD is a unicast, delay-tolerant protocol that uses beacon-driven geographic
routing to forward packets from source to destination. It adopts the idea of
carry-and-forward based on the use of predictable vehicle mobility, in order to
achieve low data delivery delay in sparse networks. Having knowledge of the
underlying road infrastructure through a static map, packets are forwarded along
streets and routing decisions take place at intersections where vehicles select
the next forwarding path (series of consecutive streets) to destination with the
smallest packet delivery delay. Through a stochastic model that takes into
consideration vehicle density on a road, road length and average vehicle
velocity, the expected packet delivery delay can be estimated.

However, the authors of~\cite{Ding07} have shown that VADD experiences
performance degradation in packet delivery delay and drop rate under
low-vehicular density conditions. Specifically, in such conditions, a packet
carrier entering an intersection, may not be able to push a packet towards the
optimal forwarding path, due to the fact that there no available vehicles on
that path. Therefore, the packet is forced to be forwarded towards a sub-optimal
path, which consequently leads to higher packet delivery delay or even higher
packet drop rate.

\subsection{GPCR: Greedy Perimeter Coordinator Routing}
GPCR is a overlay, non-delay tolerant protocol that also uses beacon-driven
geographic routing to forward packets from a source to destination. Similar with
VADD, packet forwarding is performed along streets in a greedy manner and
routing decisions are taken at intersections. GPCR takes advantage of the fact
that streets and intersections form a natural planar graph and uses it as a
repair strategy when a packet reaches a local optimum. Key to the routing
process, is the detection of nodes (coordinators) located on an intersection,
since GPCR does not rely on an underlying street map. To do so, GPCR employs two
approaches: i)~\emph{neighbor tables}, where a node \emph{x} is on an
intersection, if it has two neighbors~\emph{y} and~\emph{z} that are within
range of each other but do not list each other as neighbors, ii)
a~\emph{correlation coefficient} that relates a node with its neighbors w.r.t to
their position.

\subsection{Graph enabled VADD and GPCR}
The goal was to enable each protocol to utilize the graph analysis information,
during the routing process decisions, therefore minor changes were applied to
the proposed algorithms. Primarily, the followings changes were performed to
each aforementioned protocol - \textbf{VADD}: On each beacon period, a vehicle
broadcasts along its geo-coordinates, information about its current lobby index,
whether is a member of a cluster, and if so, the specific cluster size and
clustering coefficient. During low vehicle density conditions, such information
can be utilized by a packet carrier in a 3-step policy, as it arrives at an
intersection and its time to to choose the next packet carrier. Particularly, on
an intersection, if the current packet carrier, does not identify a candidate
node which is en route to the packet destination via an optimal path (as
calculated by MD-VADD), then in the presence of other less favorable nodes, the
next packet carrier is selected to be the node that: (1) has the largest
lobby-index value, or (2) belongs to a cluster with the largest cluster
coefficient or (3) belongs to the largest cluster in terms of membership. This
ensures that the packet will be handled to a node with higher network
connectivity, and thus increase its chances to be routed to destination.
\textbf{GPRC}: As in the VADD, on each beacon period, a vehicle broadcasts along
its geo-coordinates, information about its current lobby index value. We select
only the lobby-index to be broadcasted, since through
Section~\ref{subsec:Centrality}, we have observed that this metric can identify
vehicles that are on the close vicinity of a road intersection. In addition to
the two native algorithms for detection vehicles on an intersection, we enable
GPCR to detect such nodes using the lobby index value broadcasted on each beacon
period. In essence, the lobby-index approach is a variation of the neighbor
tables which not only identifies candidate coordinator nodes, but also provides
a view of which candidate coordinator has the best network connectivity.

\subsection{Implementation}
To achieve the above, both VADD (MD-VADD variation) and GPCR were implemented
from scratch under ns-3.11~\cite{ns-3}, trying to remain as accurate as possible
given the information provided in the original
articles~\cite{Zhao2006,Lochert2005}. Mobility traces used in network
simulations, were analyzed a priori, second-by-second, in order to extract VANET
graph information such as the lobby index values of nodes, cluster membership
and clustering co-efficient. This information was fed to ns-3 at the start of
the simulation, allowing each node, at any given time instance, to utilize it
through queries to a~\textit{God} service. It is important to notice, that how
such graph information is acquired and made available to vehicles in a real
VANET (i.e cluster detection protocols, lobby index calculation) is out of the
scope of this article.

\begin{table}[h]
\centering
\tiny
\begin{tabular}{|l|c|c|c|}\hline
& \textbf{VADD}& \textbf{GPCR} \\\hline\hline
Mobility Model & \multicolumn{2}{|c|}{IDM\_LC} \\\hline
Area Size & 4000m x 5000m & 6500m x 3500m\\\hline
Vehicles & 150 & 900 \\\hline
Vehicle Transmission Range & \multicolumn{2}{|c|}{ 300m (802.11p)} \\\hline
Propagation model & \multicolumn{2}{|c|}{Nakagami Propagation Loss} \\\hline
Simulation Time & \multicolumn{2}{|c|}{3600s (600s warm-up)} \\\hline
Beacon Interval & \multicolumn{2}{|c|}{0.5s} \\\hline
Packet Senders & 15 & 10 \\\hline
CBR Rate (pkt/sec) & 0.1 - 1 & 4 \\\hline
\end{tabular}
\caption{VADD and GPCR Simulation setup parameters}
\label{tbl:SimDetails}
\vspace{-7mm}
\end{table}

\begin{figure}[t]
    \centering
     \subfigure [VADD - Mean Packet Delivery Delay]{
        \includegraphics[width=0.32\textwidth]{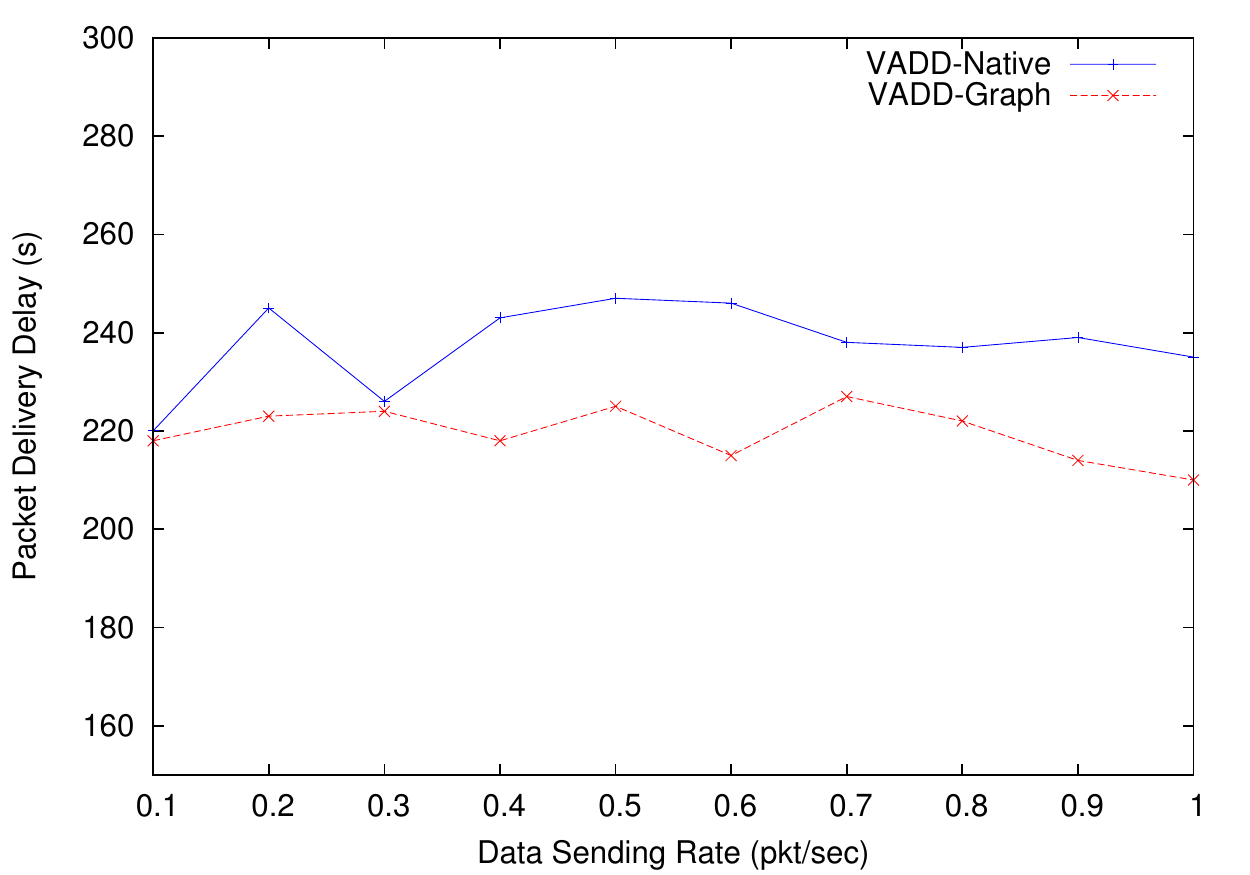}
        \label{fig:VADD}
    	}
    \subfigure [GPCR - Mean Packet Delivery Rate]{
        \includegraphics[width=0.32\textwidth]{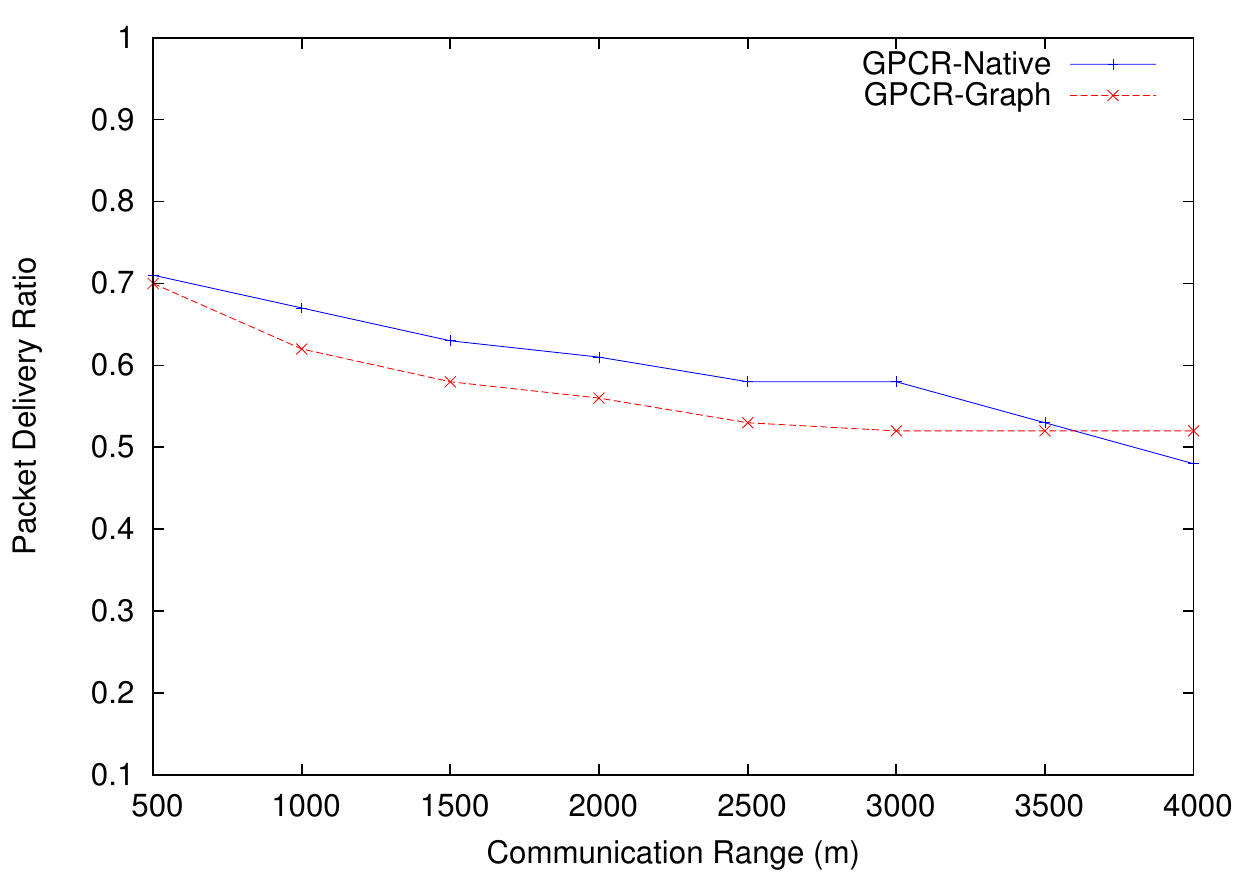}
        \label{fig:GPCR}
    }
    \caption{}
    \vspace{-6mm}
\end{figure}

Considering the modifications described above, we opted to record the packet
delivery delay over the data sending rate for VADD and packet deliver rate over
the communication distance for GPCR. Details about the simulation setup are
provided in Table~\ref{tbl:SimDetails}. Figures~\ref{fig:VADD}
and~\ref{fig:GPCR} present the results of our simulations. The figures present
the average values of each metric calculated over 5 runs of each simulation
scenario, with different random number seeds.

We observe, that VADD has a considerably high packet delivery delay when the
number of vehicle penetration is low (150 vehicles), which is consistent with the
findings of~\cite{Ding07}. By enabling VADD nodes to utilize graph information
in their routing decision on an intersection, we record an average improvement
on delivery delay of approximately 6\%, throughout the data sending rate range.
Indeed, through the ns-3 trace log we notice that when there are no nodes at the
intersection which are considered as optimal candidates by the native VADD
forwarder selection process, then lobby index information is sufficient to
identify a sub-optimal node that will be entrusted with forwarding the packet at
hand.


Concerning GPCR, we observe that by utilizing graph information, and
particularly the lobby-index value, it is sufficient to identify vehicles which
are on or at the vicinity of an intersection. Throughout the communication
range, the packet delivery rate remains above 56\%, with a maximum rate of 69\%.
It is worth to note, the utilization of such a simple graph metric for the
vehicle detection process, gives results which are slightly less better than
when utilizing the calculation of the correlation coefficient (CC).

Article~\cite{Erramilli2008} asked the question: which nodes will be the
forwarders in routing? Our study is able to provide an answer to this: we can
draw such nodes among those with high centrality value. These nodes are also
perfect candidates for message ferrying in case of network partitioning.
Similarly, nodes with high lobby index are ideal for carrying out the
rebroadcasts so as to spread the message to many recipients with as few
rebroadcasts as possible. Moreover, for applications requiring awareness of the
positions of other vehicles through periodic beacons, or the distribution of
traffic related data through periodic beacons, the exploitation of the more
''central" vehicles for these tasks could relieve the network from redundant
broadcasts and reduce the collisions.
\section{Conclusions}
\label{sec:Conclusions} 
This paper provides a complete study of the topological characteristics and
statistical features  of a VANET communication graph. Using both real and
realistic mobility traces, we study the networking shape of VANETs in urban
environments under different transmission and market penetration ranges. Given
that a number of RSUs have to be deployed for disseminating information to
vehicles in an urban area, we also study their impact on vehicular
connectivity.Specifically, our work addresses the following questions: Which are
the statistical properties that characterize the structure and behavior of
vehicular networks? How do VANET graphs evolve over time? Can be identified
communities in vehicular networks? We view our findings as particularly important
since the obtained results have a wide range of implications upon the creation of
high-performance, reliable, scalable, secure, and privacy-preserving VANET
technologies.

\small
\bibliographystyle{bib/IEEEtran}
\bibliography{bib/IEEEabrv,bib/tkde2010}

\end{document}